\documentclass[twocolumn,aps,prb,superscriptaddress]{revtex4-2}
\usepackage{color}
\usepackage{mathrsfs}
\usepackage{bm}
\usepackage{amsmath}
\usepackage{amssymb}
\usepackage{graphicx}
\usepackage{hyperref}
\usepackage{color}
\usepackage{siunitx}
\usepackage{braket}
\usepackage{tikz}
\allowdisplaybreaks[3]

\makeatletter
\hypersetup{hypertex=true, colorlinks=true, linkcolor=blue, anchorcolor=blue,citecolor=blue}

\usepackage{dcolumn}
\usepackage{bm}
\usepackage{mathrsfs}
\usepackage{amsfonts}
\usepackage{xcolor}
\usepackage{epstopdf}

\begin{document}
\title{Spin-orbital intertwined topological superconductivity in a class of correlated noncentrosymmetric materials} 
\author{Lichuan Wang}
\thanks{These authors make equal contributions.}
\affiliation{Anhui Provincial Key Laboratory of Low-Energy Quantum Materials and
		Devices, High Magnetic Field Laboratory, HFIPS, Chinese Academy
		of Sciences, Hefei, Anhui 230031, China}
\affiliation{Science Island Branch of Graduate School, University of Science and
		Technology of China, Hefei, Anhui 230026, China}
\author{Ran Wang}
\thanks{These authors make equal contributions.}
\affiliation{Anhui Provincial Key Laboratory of Low-Energy Quantum Materials and
		Devices, High Magnetic Field Laboratory, HFIPS, Chinese Academy
		of Sciences, Hefei, Anhui 230031, China}
\affiliation{Science Island Branch of Graduate School, University of Science and
		Technology of China, Hefei, Anhui 230026, China}
\author{Xinliang Huang}
\affiliation{Anhui Provincial Key Laboratory of Low-Energy Quantum Materials and
		Devices, High Magnetic Field Laboratory, HFIPS, Chinese Academy
		of Sciences, Hefei, Anhui 230031, China}
\affiliation{Science Island Branch of Graduate School, University of Science and
		Technology of China, Hefei, Anhui 230026, China}
\author{Xianxin Wu}
\email{xxwu@itp.ac.cn}
\affiliation{CAS Key Laboratory of Theoretical Physics, Institute of Theoretical Physics, Chinese Academy of Sciences,
Beijing 100190, China}
\author{Ning Hao}
\email{haon@hmfl.ac.cn}
\affiliation{Anhui Provincial Key Laboratory of Low-Energy Quantum Materials and
		Devices, High Magnetic Field Laboratory, HFIPS, Chinese Academy
		of Sciences, Hefei, Anhui 230031, China}
\affiliation{Science Island Branch of Graduate School, University of Science and
		Technology of China, Hefei, Anhui 230026, China}
	
	\begin{abstract} 
		In this study, we propose an alternative route to realizing topological superconductivity (TSC). Our approach targets a new class of correlated noncentrosymmetric materials that host two spin-split Fermi surfaces due to spin-orbital intertwined effects. By investigating the superconducting pairings based on two-orbital Rashba-Hubbard model on a square lattice within a spin-fluctuation-mediated pairing framework, we find that, depending on model parameters, the leading superconducting state belongs to the $A_1(S_{\pm})$, $B_2$ or $B_2(d_{\pm})$ irreducible representations (IRs) of the $C_{4v}$ point group. Notably, the $A_1(S_{\pm})$ state features a sign-changed gap between the two spin-split Fermi surfaces and exhibits two key characteristics: (i)  it is a parity-mixed state, with the ratio of even-parity to odd-parity components tunable via  the Rashba spin-orbit coupling (RSOC) strength and onsite orbital-dependent potential; (ii) it is a fully gapped TSC, characterized by a $Z_2$ topological invariant. When RSOC is strong, the $B_2(d_{\pm})$ state can emerge and exhibits both parity-mixed and nodal TSC features. Further analysis reveals that the fully gapped TSC can be predominated by spin-singlet, despite of the presence of the spin-triplet components. This distinguishes our case from single-orbital noncentrosymmetric materials, where TSC arises from $p$-wave or $f$-wave spin-triplet pairing. In those systems, the Fermi level must be tuned near a van Hove singularity to enhance ferromagnetic fluctuations that drive the spin-triplet pairing. These distinctions enhance the experimental feasibility of our model.
	\end{abstract}
	\maketitle
	
	\section{Introduction} TSCs can host Majorana bound states, which are exotic quasi-particles obeying non-Abelian statistics. This unique feature makes TSCs fundamentally fascinating and potentially valuable for applications in quantum computing\cite{ivanov2001non, kitaev20031, kitaev2006anyons, nayak2008non, alicea2012new, elliott2015colloquium, RevModPhys.82.3045, RevModPhys.83.1057, PRB-1}. Over the past two decades, many efforts have been devoted to realizing TSCs in various material platforms\cite{fu2008superconducting, sau2010generic, nadj2014observation, hao2014topological, wang2015topological, wu2016topological,   hao2019topological, wu2015cafeas, xu2016topological}. The general strategy involves utilizing superconducting proximity effects in either momentum space or real space. The former, also known as connate TSCs, is typically achieved in multiband superconducting systems with topological boundary states, such as Fe(Se,Te), PbTaSe$_{2}$ etc\cite{ hao2014topological, wang2015topological, wu2016topological,   hao2019topological, wu2015cafeas, xu2016topological, chang2016topological, guan2016superconducting}. The latter approach involves the artificial construction of topological insulator (or semiconductor)-superconductor heterostructures\cite{fu2008superconducting, sau2010generic, nadj2014observation}. However, these material platforms pose experimental challenges, requiring high interface quality, a large induced superconducting gap, and precise control of the chemical potential.

Unlike proximity-effect-induced TSCs, intrinsic TSCs exhibit topological properties throughout their entire volume, making them robust against perturbations at boundaries or interfaces\cite{ivanov2001non}. This characteristic provides a significant advantage, positioning intrinsic TSCs as strong candidates for building quantum computing platforms.
The pairing symmetry in intrinsic TSCs is typically $p$-wave spin-triplet, which generally arises only in unconventional superconducting systems driven by non-electron-phonon coupling mechanisms, making such materials exceedingly rare in nature. Current candidates include uranium- and cerium-based heavy fermion compounds, such as UTe$_2$, CePt$_3$Si\cite{jiao2020chiral, aoki2019unconventional, aoki2022unconventional, bauer2004heavy, smidman2017}. Consequently, the exploration of new material systems for realizing TSCs remains an important and active area of research.     

A comprehensive analysis of material systems exhibiting TSC reveals that their Fermi surfaces typically form robust spin textures, indicating explicit or potential topological characteristics in their electronic structures. Notably, materials such as CePt$_3$Si, Y$_2$C$_3$, A$_2$Cr$_3$As$_3$ (A=K, Rb, Cs)\cite{bauer2004heavy, smidman2017, krupka1969high, bao2015superconductivity, tang2015unconventional, tang2015superconductivity}, which lack central inversion symmetry, display similar features due to RSOC. Comparable spin texture characteristics have also been experimentally observed in the Fermi surfaces of certain copper- and iron-based superconductors\cite{gotlieb2018revealing, borisenko2016}. These insights suggest that TSCs may emerge in correlated unconventional superconducting systems when RSOC is considered. However, current research on single-orbital Rashba-Hubbard models indicates that the small contribution of the spin-triplet component in these systems is typically insufficient to realize TSCs, except under specific conditions, such as fillings close to the van Hove singularity\cite{greco2018mechanism, nogaki2020, greco2020, bonetti2024}.

To address the challenges in realizing intrinsic TSCs, we propose an alternative approach by introducing a new degree of freedom, such as orbital, layer, or valley. For clarity, we focus on orbital degrees of freedom and consider a two-orbital Rashba model, in which the two spin-split Fermi surfaces can exhibit either conventional or unconventional spin textures. This transition can be tuned by an on-site orbital-dependent potential together with onsite SOC, reflecting the evolution from the single-orbital limit to the two-orbital limit. Using the random phase approximation (RPA), we solve the superconducting problem of this model within the framework of spin-fluctuation-mediated superconductivity. We find that, depending on the strength of onsite SOC and effective RSOC, onsite orbital-dependent potential and filling level, the Hubbard interaction can drive the leading pairing symmetry into the $A_1(S_{\pm})$, $B_2$ or $B_2(d_{\pm})$ IRs of the $C_{4v}$ point group. Notably, the $A_1(S_{\pm})$ IR is conformed to host an $S_{\pm}$-wave pairing state, which lead to a fully gapped TSC characterized by a $Z_2$ topological invariant. Furthermore, the $S_{\pm}$-wave TSC is a parity-mixed state, with  the parity-mixed ratio tunable via the strength of onsite SOC and effective RSOC and onsite orbital-dependent potential. This tunability offers a promising platform for exploring parity-mixed superconducting states. More significantly, the TSC can be predominatntly driven by spin-singlet component, despite of the spin-triplet contribution. We also find the $B_2(d_{\pm})$ state can emerge under the strong RSOC and exhibits both parity-mixed and nodal TSC features.

\section{Model and method} 
\begin{figure}[h]
	\centering
	\includegraphics[width=1.0\columnwidth]{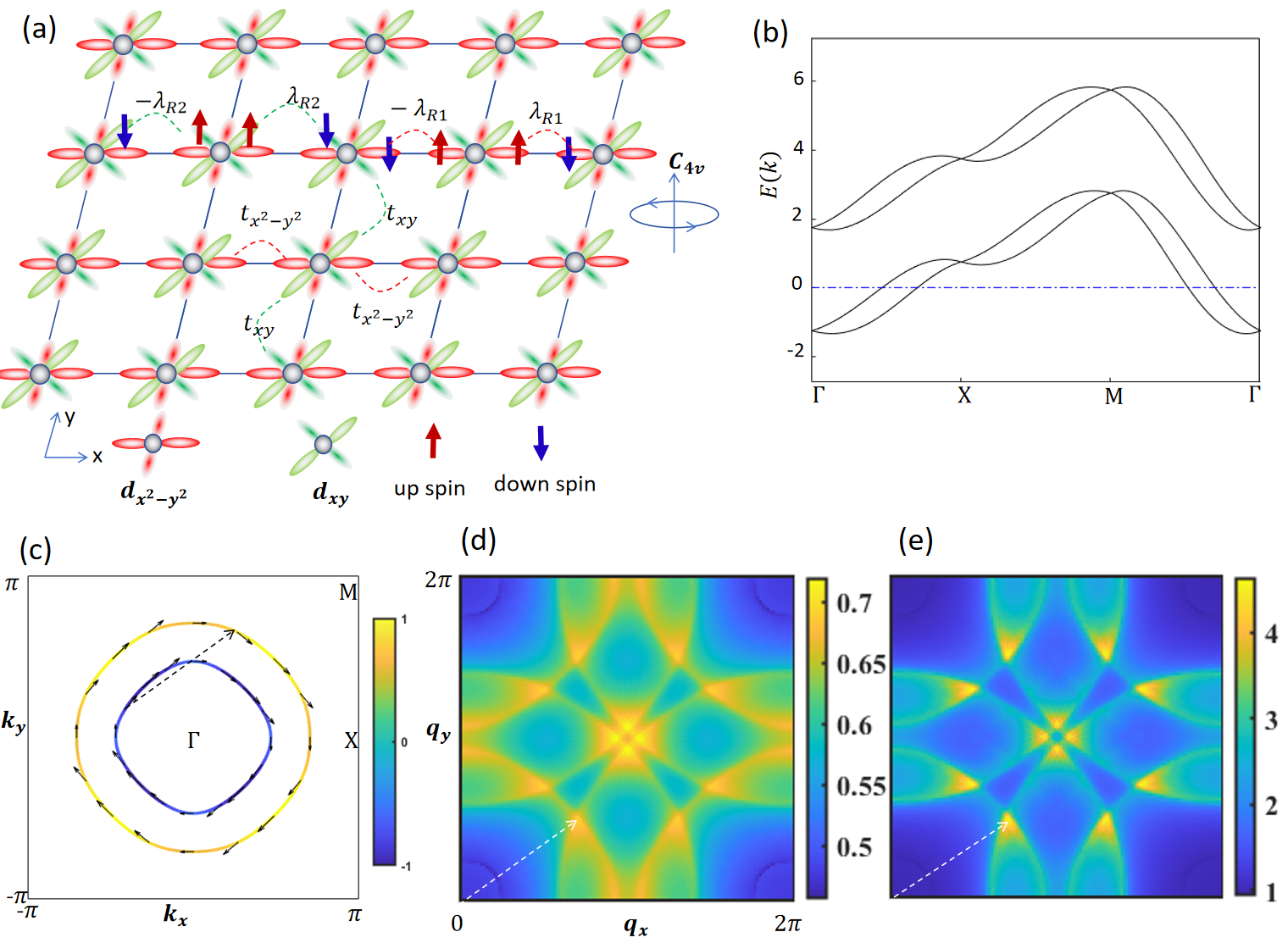}
	\caption{(a) The two-orbital ($d_{x^2-y^2},d_{xy}$) unconventional Rashba model defined on a square lattice with $C_{4v}$ symmetry. The polar axis is defined as the axis perpendicular to the square lattice plane, indicating that along the two opposite directions of this axis the system experiences different polarization fields $V(z)\neq V(-z)$. This asymmetry can in turn generate Rashba-type SOC. (b) The transition from single-orbital limit with conventional spin texture (left part in (b)) to two-orbital limit with unconventional spin texture (right part in (b)) controlled by $\epsilon_{0}^{onsite}$ or $\lambda_{so}$. The circles with cyan color denote the Fermi surfaces and the red arrows indicate the spin texture. (c) The band structures under different values of parameters 
	$\epsilon_{0}^{onsite}$ or $\lambda_{so}$. From left to right,  the parameters are set to be $\lambda_{so}=0$, $\lambda_{so}=1.5$ and $\lambda_{so}=3$ respectively. $\lambda_{so}=3+\epsilon_{0}^{onsite}$ is also fixed in order to prevent significant changes in the band structure. Other parameters are set as follows: $t^{nn}_2=-0.1$, $\lambda_R=0.5$, $\epsilon=1$, $t^{nnn}_1=0.0125$, $t^{nnn}_2=0.0375$, the occupied particle number per site is set to be $n=0.5$.  Here, the  coordinates of high-symmetry points are defined as follows: $\Gamma=(0,0)$; $X=(\pm\pi,0)$, $(0,\pm\pi)$; $M=(\pm\pi,\pm\pi)$.  }
	\label{band}
\end{figure}

The two-orbital Rashba-Hubbard model\cite{huang2024, wang2024, wang2025finite, huang2024-1} is given by
		\begin{equation}
		H=H_0+H_{int},
\label{Htot}
	\end{equation}
with
	\begin{equation}
		H_0=\sum_{k} \xi (k)-\lambda_R (\tau^0+\epsilon\tau^3) [\bm{\sigma} \times \bm{\nabla_k} \gamma_1(k)]_{z}+\lambda_{so} \tau^{2}\sigma^3,
\label{H0}
	\end{equation}
in which, the SOC terms have the similar forms of those in the Kane-Mele model \cite{PhysRevLett.95.146802}.RSOC is origined from $H_{RSOC}\propto (\nabla V \times p)\cdot \sigma$ under the condition $V(z)\neq V(-z)$ with $V(\pm z)$being the polarization fields along the ploar axis defined in Fig. \ref{band} (a).
Here, $H_0$ can be defined on a square lattice with $C_{4v}$ symmetry, as shown in Fig. \ref{band} (a) . The two orbitals are fixed to be $d_{x^2-y^2}$ and $d_{xy}$. Then,  $\xi (k)=2(t^{nn}_1\tau^0+t^{nn}_2\tau^3)\gamma_1(k)+4(t^{nnn}_1\tau^0+t^{nnn}_2\tau^3)\gamma_2(k)-\epsilon_{0}^{onsite}\tau^3-\mu$ with $\gamma_1(k)=\cos k_x+\cos k_y$, $\gamma_2(k)=\cos k_x\cos k_y$, $\epsilon_{0}^{onsite}$ indicating the difference of on-site orbital-dependent potential and $\mu$ denoting the chemical potential. $\lambda_R$ represents the strength of effective RSOC, and a parameter $\epsilon$ is introduced to measure the RSOC anisotropy of two orbitals\cite{note1}. $\lambda_{so}$ is the strength of on-site SOC. To match the parameter settings in Fig. \ref{band} (a), $t^{nn/nnn}_{1/2}=(t^{nn/nnn}_{x^2-y^2}\pm t^{nn/nnn}_{xy})/2$ with $t^{nn/nnn}_{x^2-y^2/xy}$ denoting the nearest (nn) and next-nearest (nnn) neighbor hoppings for two $d_{x^2-y^2/xy}$ orbitals  and $\lambda_R=(\lambda_{R1}+\lambda_{R2})/2$. In the following, $t^{nn}_1$ is set to $-1$, and other parameters are in unit of $|t^{nn}_1|$. $\lambda_{so}$ is set to $3+\epsilon_{0}^{onsite}$ throughout the calculations. Pauli matrices $(\tau^0, \bm{\tau})$ and $(\sigma^0, \bm{\sigma})$ span in orbital and spin spaces, respectively. Then, $H_0$ spans in the basis $\Psi_k = (c_{x^2-y^2,k,\uparrow} , c_{x^2-y^2k,\downarrow}, c_{xy,k,\uparrow}, c_{xy,k,\downarrow})^T$ with $c_{\alpha k,\sigma}$ being the electron annihilation operator. 

The Hamiltonian $H_0$ in Eq. (\ref{H0}) yields the representative band structure, Fermi surfaces, and spin textures shown in Fig. \ref{band}. We find that when the parameter $\epsilon_{0}^{onsite}<-0.46$ or $\lambda_{so}<2.54$, the two spin-split Fermi surfaces display a conventional spin texture, indicating that the model reduces to the single-orbital limit. In contrast, when $\epsilon_{0}^{onsite}>-0.46$ or $\lambda_{so}>2.54$, the two spin-split Fermi surfaces exhibit an unconventional spin texture, signaling that the model is in the two-orbital limit. This property will impact the magnitude of transverse susceptibility and ultimately affect the effective inter-pocket interaction. The below discussions will cover both limits.

The two-orbital Hubbard interactions can be expressed as: 
\begin{equation}
	\begin{aligned}
		H_{int}&=U\sum_{i \alpha}\hat{n}_{i \alpha\uparrow}\hat{n}_{i \alpha\downarrow}
		+V\sum_{i \alpha<\beta}\sum_{\sigma\sigma'}\hat{n}_{i \alpha \sigma}\hat{n}_{i \beta \sigma'}\\
		&-J\sum_{i, \alpha<\beta}\bm{S}_{i\alpha}\cdot\bm{S}_{i\beta}\\
		&+J'\sum_{i, \alpha<\beta}\sum_{\sigma}c^{\dagger}_{i\alpha\sigma}c^{\dagger}_{i\alpha\bar{\sigma}}
		c_{i\beta\bar{\sigma}}c_{i\beta\sigma}.
	\end{aligned}
\label{Hint}
\end{equation}
 Here, $U$, $V$, $J$ and $J'$ represent the intra- and inter-orbital on-site repulsive Hubbard interactions, Hund's coupling, and the pairing hopping term, respectively. $\hat{n}_{i \alpha \sigma}$ and $\bm{S}_{i\alpha}$ denote the spin-$\sigma$ electron density operator and spin operator for electrons in orbital $\alpha$ on $i$ site, respectively. For clarity, we adopt the conventional notations $J=J'$ and $V=U-2J$ in the following discussion. 
\begin{figure*}[!htbp]
	\centering
	\includegraphics[width=7in]{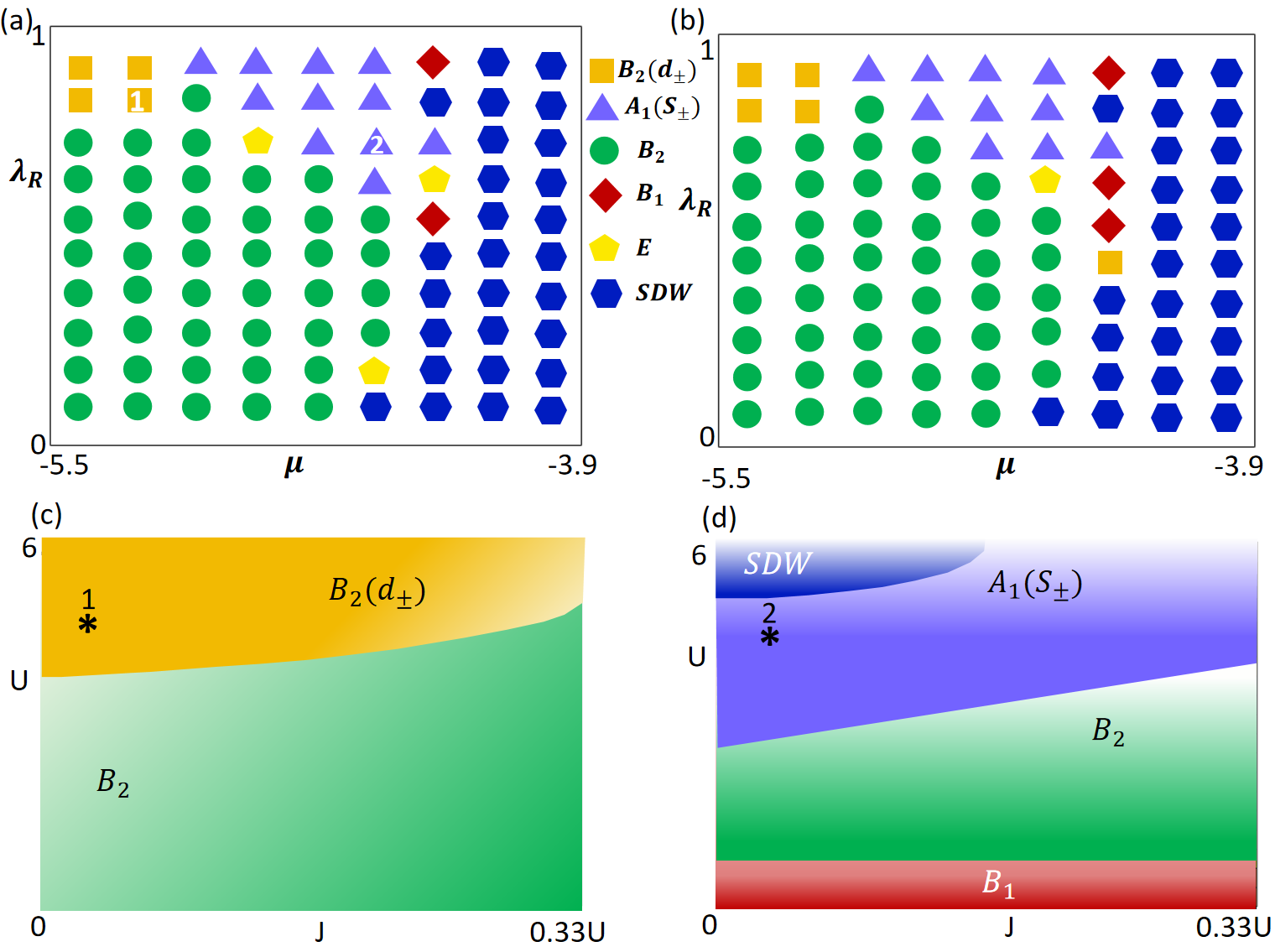}
	\caption{ (a), (f), (k) and (p) The plottings illustrate Fermi surfaces under various parameters: for (a),$\lambda_{so}=0$, $n=0.5$, $U=3.5$ and $\lambda_{R}=0.35$; for (f), $\lambda_{so}=3$, $n=0.5$, $U=3.5$and $\lambda_{R}=0.725$; for (k), $\lambda_{so}=0$, $n=0.28$, $U=7.5$and $\lambda_{R}=0.225$; for (p), $\lambda_{so}=3$, $n=0.26$, $U=7.5$and $\lambda_{R}=0.35$. The red and blue colors of the Fermi surfaces indicates the superconducting gap functions have opposite signs. Additionally, black arrows denote the spin texture of the relevant Fermi surface on the $k_x-k_y$ plane, while red arrows indicates a $\bm{q}$-vector corresponding to the peaks of $\chi^{RPA}_{zz}(\bm{q})$. The points on the Fermi surface are labeled from 1 to 64 for the outer Fermi surface and from 65 to 128 for the inner Fermi surface in (a), and the settings are the same in (f), (k) and (p). Figures (b), (g), (l) and (q) display $\chi^{RPA}_{zz}(\bm{q})$ defined in Eq. \ref{sus_sca} corresponding to (a), (f), (k) and (p) respectively, whereas figures (c), (h), (m) and (r) show the relevant $\chi^{RPA}_{+-}(\bm{q})$ defined in Eq. \ref{sus_sca}. Figures (d), (i), (n) and (s) present the intra-pocket effective interaction in $\tau^0\sigma^0$ channel of the outer Fermi surface corresponding to figures (a), (f), (k) and (p), while figures (e), (j), (o) and (t) depict the inter-pocket effective interaction in $\tau^0\sigma^0$ channel between different Fermi surfaces. The effective intra-pocket interaction of the inner Fermi surface can be neglected, because it is weaker than the other two types of interactions, as indicated by the absence of peaks connecting the inner Fermi surface to itself in $\chi^{RPA}_{zz}(\bm{q})$ in (l) and (q). On the right, for figures (d), (e), (i) and (j) correspond to the upper color bar, while for figures (n), (o), (s) and (t) correspond to the lower color bar.}
	\label{fs}
\end{figure*}
The general bare susceptibility can be expressed as follows,
\begin{equation}
	\begin{aligned}
		&\chi^{l_1\sigma_1l_2\sigma_2}_{l_3\sigma_3l_4\sigma_4}(\bm{q},i\nu_n)&\\
		&=-T\sum_{\bm{k},i\omega_n}G_{l_3\sigma_3l_1\sigma_1}(\bm{k},i\omega_n)G_{l_2\sigma_2l_4\sigma_4}(\bm{k},i\omega_n-i\nu_n),
	\end{aligned}
\label{bare}
\end{equation}
with the Matsubara Green function,
\begin{equation}
	\begin{aligned}
		\hat{G}(\bm{k},i\omega_n)=[i\omega_n-H_0(\bm{k})]^{-1}.
	\end{aligned}
\label{greenfunc}
\end{equation}
Here, $l_{1-4}$ are orbital index, and $\sigma_{1-4}$ are spin index. $i\omega_{n}$ and $i\nu_n$ fermionic and bosonic Matsubara frequencies, respectively. Using RPA, the dressed susceptibility can be calculated by
\begin{equation}
	\begin{aligned}
\hat{\chi}^{RPA}(\bm{q},i\nu_n)=[\hat{1}-\hat{\chi}(\bm{q},i\nu_n)\hat{W}]^{-1}\bm{\chi}(\bm{q},i\nu_n).
	\end{aligned}
\label{RPAsus}
\end{equation}
Here, $\hat{W}$ is the bare interaction matrix with non-zero elements $W^{l_1\sigma l_1\bar{\sigma}}_{l_1\sigma l_1\bar{\sigma}}=U$, $W^{l_1\sigma l_1\sigma}_{l_1\bar{\sigma} l_1\bar{\sigma}}=-U$, $W^{l_2\sigma l_1\bar{\sigma}}_{l_2\sigma l_1\bar{\sigma}}=V$, $W^{l_1\sigma l_1\sigma}_{l_2\bar{\sigma} l_2\bar{\sigma}}=-V$, $W^{l_1\sigma l_1\bar{\sigma}}_{l_2\sigma l_2\bar{\sigma}}=J$, $W^{l_1\sigma l_2\sigma}_{l_1\bar{\sigma} l_2\bar{\sigma}}=-J$, $W^{l_1\sigma l_2\bar{\sigma}}_{l_1\sigma l_2\bar{\sigma}}=J'$, and $W^{l_1\sigma l_2\sigma}_{l_2\bar{\sigma} l_1\bar{\sigma}}=-J'$.

According to the theory of fluctuation-mediated superconductivity\cite{berk1966, emery1964, scalapino2012}, the effective static superconducting pairing interactions can be calculated by considering bubble and ladder diagrams shown in Fig. \ref{Feynman}, and are expressed as:
\begin{equation}
	V^{l_1\sigma_1l_2\sigma_2}_{l_3\sigma_3l_4\sigma_4}(\bm{k},\bm{k'})=W^{l_1\sigma_1l_2\sigma_2}_{l_3\sigma_3l_4\sigma_4}+V_{l}(\bm{k},\bm{k'})-V_{b}(\bm{k},\bm{k'}),
	\label{effv}
\end{equation}
with the ladder term
\begin{equation}
	V_{l}(\bm{k},\bm{k'})=[\hat{W}\hat{\chi}^{RPA}\hat{W}]^{l_1\sigma_1l_2\sigma_2}_{l_3\sigma_3l_4\sigma_4}(\bm{k}+\bm{k'}),
	\label{effv1}
\end{equation}
and the bubble term
\begin{equation}
	V_{b}(\bm{k},\bm{k'})=[\hat{W}\hat{\chi}^{RPA}\hat{W}]^{l_1\sigma_1l_4\sigma_4}_{l_3\sigma_3l_2\sigma_2}(\bm{k}-\bm{k'}).
	\label{effv2}
\end{equation}
The possible superconducting pairing can be evaluated through solving the linear Eliashaberg equation as follows,
\begin{equation}
\begin{aligned}
&\lambda\Delta_{l_1\sigma_1l_4\sigma_4}(\bm{k})&\\
&=T\sum_{\bm{k'}i\omega_n}\sum_{l_2l_3\sigma_2\sigma_3}V^{l_1\sigma_1l_2\sigma_2}_{l_3\sigma_3l_4\sigma_4}(\bm{k},\bm{k'})F_{l_3\sigma_3l_2\sigma_2}(\bm{k'},i\omega_n),
\end{aligned}
	\label{gap_func}
\end{equation}
with
\begin{equation}
\begin{aligned}
&F_{l_3\sigma_3l_2\sigma_2}(\bm{k'},i\omega_n)&\\
&=G_{l_3\sigma_3l\sigma}(\bm{k'},i\omega_n)\Delta_{l\sigma l'\sigma'}(\bm{k'})G_{l_2\sigma_2l'\sigma'}(-\bm{k'},-i\omega_n).
\end{aligned}
	\label{gap_func1}
\end{equation}
Here, $\Delta_{l_1\sigma_1l_4\sigma_4}(\bm{k})$ is the superconducting pairing function and the leading superconducting instability can be identified by the largest positive eigen-value of $\lambda$. More details can be found in Appdenix A.

\section{Fermionology and spin susceptibility}
Previous studies on the conventional Rashba-Hubbard model have focused on cases where the filling is near the van Hove singularity. This condition induces strong ferromagnetic fluctuations, which, in turn, favor the emergence of leading $p$- or $f$-wave spin-triplet superconducting states\cite{greco2018mechanism, greco2020, bonetti2024}. In this work, we relax this restriction and consider more general cases of electron doping beyond the van Hove singularity. A typical band structure and Fermi surfaces under electron doping are shown in the Figs. \ref{band} (c) and Figs. \ref{fs}(a), (f) respectively.

To analyze the behavior of the spin  susceptibilities, we define the scalar forms of longitudinal and transverse spin susceptibilities $\chi^{RPA}_{zz}(\bm{q})$ and $\chi^{RPA}_{+-}(\bm{q})$ as follows,
\begin{equation}
\begin{aligned}
&\chi^{RPA}_{zz}(\bm{q})=\sum_{l_1,l_2}\sum_{\{\sigma\}}\sigma^z_{\sigma_1\sigma_2}\sigma^z_{\sigma_4\sigma_3}\chi^{l_1\sigma_1l_1\sigma_2}_{l_2\sigma_3l_2\sigma_4}(\bm{q}),&\\
&\chi^{RPA}_{+-}(\bm{q})=\sum_{l_1,l_2}\sum_{\{\sigma\}}\sigma^+_{\sigma_1\sigma_2}\sigma^{-}_{\sigma_4\sigma_3}\chi^{l_1\sigma_1l_1\sigma_2}_{l_2\sigma_3l_2\sigma_4}(\bm{q}). 
\end{aligned}
	\label{sus_sca}
\end{equation}
	The RPA spin susceptibilities corresponding to the Fermi surfaces shown in Fig. \ref{fs} (a), (f), (k) and (p) can be calculated using Eqs. \ref{bare}, \ref{RPAsus} and \ref{sus_sca}. The difference between longitudinal and transverse susceptibility originates from the breaking of spin-rotation invariance induced by spin-splitting effects such as RSOC here. Specifically, in the single-orbital limit, the longitudinal susceptibility, as illustrated in Fig. \ref{fs}(l), can be simplified as
\begin{equation}
		\chi^{RPA}_{zz}(\bm{q})=\sum_{\mu\nu,k\sigma,l_1l_2}C_{l_1,l_2}^{\mu \nu}[1-\cos (\theta_k-\theta_{k+q})]\times g[E],
		\label{sus-b1}
\end{equation}
where $e^{i\theta_k}=(\sin k_y +i\sin k_x)/\sqrt{\sin^2k_y+\sin^2k_x}$, 
 $g[E]=\frac{f(E_{\nu}(\bm{k}+\bm{q}))-f(E_{\mu}(\bm{k}))}{E_{\nu}(\bm{k}+\bm{q})-E_{\mu}(\bm{k})+i0^+}$ 
 and the coefficient $C_{l_1,l_2}^{\mu \nu}$ depend on $l_1, l_2, \mu, \nu$ and $k$  with $\mu$, $\nu$ denoting the band indices. For intra-pocket crossings, where $k=-(k+q)$, one has $\theta_k=\theta_{k+q}+\pi$, and therefore $1-\cos (\theta_k-\theta_{k+q})=2$, so that the susceptibility reaches its maximum. In contrast, for inter-pocket crossings this condition is not satisfied, and the susceptibility does not reach its maximum value. This explains why two distinct types of peaks are observed in Fig. \ref{fs}(l).
Meanwhile, the transverse susceptibility, shown in Fig. \ref{fs}(m), can be written as
	\begin{equation}
		\chi^{RPA}_{+-}(\bm{q})=\sum_{\mu\nu,k}|a^{\uparrow}_{\mu}(k)|^2|a^{\downarrow}_{\nu}(k+q)|^2\times g[E].
		\label{sus-b2}
	\end{equation}
The wave-function factor $|a^{\uparrow}_{\mu}(k)|^2|a^{\downarrow}_{\nu}(k+q)|^2$ is nearly constant (approximately 0.25) for both intra-pocket and inter-pocket crossings. As a result, three types of peaks appear, corresponding to intra-pocket crossing of the outer Fermi surface, inter-pocket crossing between different Fermi surfaces, and intra-pocket crossing of the inner Fermi surface.

	Now, we focus on $\chi^{RPA}_{zz}(\bm{q})$ as illustrated in Figs. \ref{fs} (b), (g), (l) and (q), which is significantly larger than the $\chi^{RPA}_{+-}(\bm{q})$, as shown in Figs. \ref{fs} (c), (h), (m) and (r). Note that owing to the $C_{4v}$ symmetry of the system, all RPA spin susceptibility patterns shown in Fig. \ref{fs} manifest a characteristic fourfold rotational symmetry. $\chi^{RPA}_{zz}(\bm{q})$ primarily governs the characteristics of spin fluctuations within the system. For $n=0.5$, we set $U=3.5$ to ensure that the eigenvalue of the gap function approaches 1. The peaks of $\chi^{RPA}_{zz}(\bm{q})$ correspond to the intersection of the outer Fermi surface's 2$k_{F1}$ circle and the inner Fermi surface's 2$k_{F2}$ circle, approximately at $\bm{q}=(0.75\pi, 0.44\pi)$. Notably, this $\bm{q}$ coincides with the inter-pocket connecting vector illustrated in Fig. \ref{fs} (a). This is referred to as ``double nesting'' as previously researched \cite{PhysRevB.81.014509} . Consequently, for superconducting pairing driven by Hubbard $U$ and mediated by spin fluctuations (as shown in Fig. \ref{fs} (b)), the Fermi surfaces represented in Fig. \ref{fs} (a) tend to favor a sign-changed gap function—namely, the superconducting gap functions on the two Fermi surface pieces connected by vector $\bm{q}$ adopt opposite signs. This picture also applies to the two-orbital limit shown in Figs. \ref{fs} (f) and (g).  When $n\approx0.28$, we set $U=7.5$ to again ensure the eigenvalue of the gap function approaches 1. The peaks of $\chi^{RPA}_{zz}(\bm{q})$ correspond to the maxima of the outer Fermi surface's 2$k_{F1}$ circle and the inner Fermi surface's $2k_{F2}$ circle, which is also coincides with the inter-pocket connecting vector illustrated in Fig. \ref{fs} (k), as depicted in Fig. \ref{fs} (l). This picture also applies to the two-orbital limit shown in Figs. \ref{fs} (p) and (q). These two types of peaks compete with one another, utimately determine the superconducting gap functions.
	Due to the settings of parameters, the band structure remains relatively unchanged while $\lambda_{so}$ varies, resulting in the longitudinal susceptibility remaining between Figs. \ref{fs} (b) and (g), as well as (l) and (q) leading to similar physical results. This indicates that the longitudinal susceptibility in Figs. \ref{fs} (b), (g), (l) and (q) play a crucial role in determining the characteristics of the potential superconducting state.
 A typical sign distribution of the gap function with $S_{\pm}$-wave pairing is indicated by the red and blue colors in Fig. \ref{fs} (a), (f), (k) and (p).

	We then examine the impact of transverse $\chi^{RPA}_{+-}(\bm{q})$. The numerical results indicate that for $\lambda_{so}=3$ (two-orbital limit) as shown in Fig. \ref{fs} (h) and (r), $\chi^{RPA}_{+-}(\bm{q})$ is approximately zero, leading to a comparable strength of effective interactions between intra-pocket and inter-pocket interactions, as shown in Fig. \ref{fs} (i) and (j) or (s) and (t). A stronger inter-pocket interaction favors a inter-pocket sign-change gap function  making $\tau^0\sigma^0$ channel more likely to contribute to this sign change. Conversely, for $\lambda_{so}=0$ (one-orbital limit), the relatively strong $\chi^{RPA}_{+-}(\bm{q})$ depicted in Fig. \ref{fs} (c) and (m) renders the effective inter-pocket interaction being significantly weaker than the intra-pocket interaction as shown in Fig. \ref{fs} (d) and (e) or (n) and (o), thereby reducing the contribution of the $\tau^0\sigma^0$ channel to the inter-pocket sign-change behavior in the gap function. This indicates that the $\chi^{RPA}_{+-}(\bm{q})$ in Fig. \ref{fs} (c), (h), (m) and (r) plays a important role in influencing the contributions from different interaction channels, resulting in the  $\tau^0\sigma^0$ channel of the two-orbital model being more significant and stable than that of the model in single-orbital limit, thereby distinguishing it from the latter.

\section{Parity-mixed state and $S_{\pm}$-wave state} Once superconductivity is induced by antiferromagnetic fluctuations, the pairing symmetry of the superconducting state can be classified according to the IRs of the $C_{4v}$ group. Table \ref{pair-class} lists the possible pairings with their form factors up to the next-nearest neighbor and their corresponding IRs.
\begin{table}
	\caption{Possible IRs of superconducting pairings under the constraints of group $C_{4v}$. Here, $g_k=\sin k_x\sin k_y$.} 
	\label{pair-class}
	\begin{ruledtabular}
\begin{tabular}{ccc}
Pairing form & $\phi _{g/u}$ / $\boldsymbol{d_{g/u}}$ & IRs \\ \hline
$i\phi _{g}\sigma ^{2}\tau ^{0/3(1)}$ & 1, $\cos k_{x}+\cos k_{y}$, $\cos k_{x}\cos k_{y}$ & $%
A_{1}(A_{2})$ \\ 
& $\cos k_{x}-\cos k_{y}$ & $B_{1}(B_{2})$ \\ 
& $\sin k_{x}\sin k_{y}$ & $B_{2}(B_{1})$ \\ 
$i\phi _{u}\sigma ^{2}\tau ^{2}$ & $\{\sin k_{x},\sin k_{y}\}$ & $E$ \\ 
$i(\bm{d_u}\cdot \bm{\sigma})\sigma ^{2}\tau ^{0/3(1)}$ & $(-\sin k_{y},\sin
k_{x},0)$ & $A_{1}(A_{2})$ \\ 
& $(\sin k_{x},\sin k_{y},0)$ & $A_{2}(A_{1})$ \\ 
& $(\sin k_{y},\sin k_{x},0)$ & $B_{1}(B_{2})$ \\ 
& $(\sin k_{x},-\sin k_{y},0)$ & $B_{2}(B_{1})$ \\ 
& $[(0,0,\sin k_{x}),(0,0,\sin k_{y})]$ & $E$ \\ 
$i(\bm{d_g}\cdot \bm{\sigma})\sigma ^{2}\tau ^{2}$ & $(0,0,\cos k_{x}+\cos
k_{y})$ & $A_{1}$ \\ 
& $(0,0,\cos k_{x}-\cos k_{y})$ & $B_{1}$ \\ 
& $(0,0,g_k)$ & $B_{2}$ \\ 
& $\{(\cos k_{x/y},0,0),(0,\cos k_{y/x},0)\}$ & $E$ \\ 
& $\{(g_k,0,0),(0,g_k,0)\}$ & 
\end{tabular}
	\end{ruledtabular}
\end{table}

\begin{figure}
	\centering
	\includegraphics[width=1.0\columnwidth]{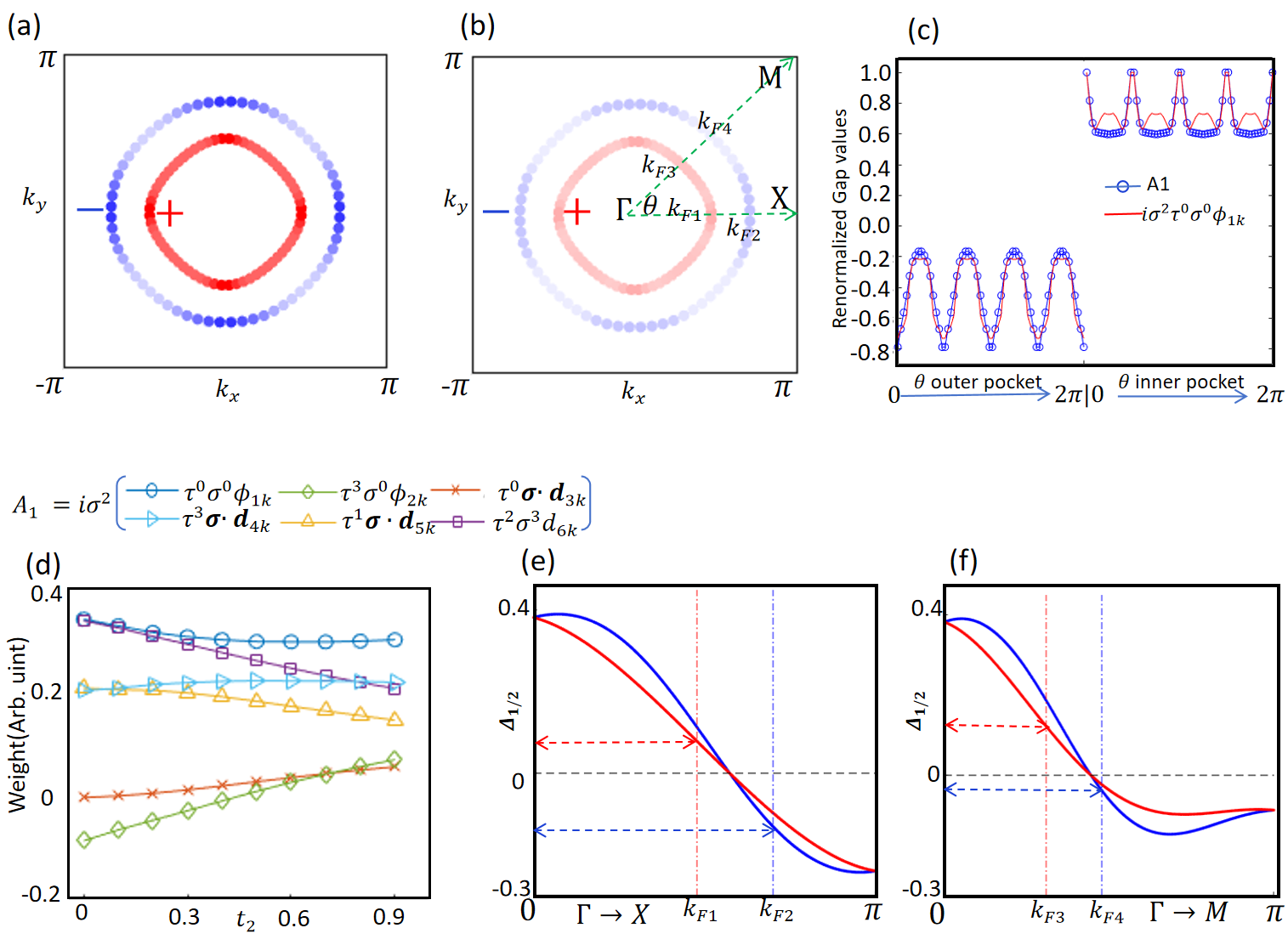}
	\caption{  (a), (b), (c) and (d) present the phase diagram under varuios parameters. In (a), $\lambda_{so}=0$ and $U=3.5$; in (b), $\lambda_{so}=3$ and $U=3.5$; in (c),$\lambda_{so}=0$ and $U=7.5$; in (d), $\lambda_{so}=3$ and $U=7.5$. The x-axis represents particle number, varying from 0.2 to 0.5 for $U=3.5$ and from 0.2 to 0.34 for $U=7.5$. The y-axis denotes the strength of RSOC $\lambda_{R}$, varying from 0.1 to 0.55 for $\lambda_{so}=0$ and from 0.1 to 1 for $\lambda_{so}=3$. Cyan points indicate a sign-changing $s$-wave superconducting gap, with $\tau^0\sigma^0$ as the leading channel contributing most to the gap. Dark blue points represent a similar type of gap, except triplet channels as the leading contributors. Yellow points denote an ordinary $d$-wave gap, while red points signify a $d_{\pm}$-wave gap, where the gap exhibits opposite signs on different Fermi surfaces. Green points represent SDW states, characterized by RPA suceptibility that tends toward divergence.
}
	\label{phase}
\end{figure}

By solving Eq. (\ref{gap_func}), the leading pairing can be determined.
The phase diagrams concerning $\lambda_R-n$ with $n$ being the occupied particle number per site under various parameters are presented in Fig. \ref{phase}. Across a wide-range parameter space, we identify two possible leading pairing channels: $A_1(S_{\pm})$, and $B_2$ as shown in Figs. \ref{phase} (a) and (b). Notably, the $A_1(S_{\pm})$ state exhibits a highly unconventional sign-change gap function, i.e., $S_{\pm}$-wave pairing, as shown in Figs. \ref{fs} (a), (f), (k), (p). The phase diagrams in Fig. \ref{phase} indicate that when the strength of RSOC is weak, the superconducting gap maintains a conventional $d$-wave gap, consistent with numerical results from the single-orbital model of cuprates. When the Hubbard $U$ is relatively small, a strong RSOC, leads to a inter-pocket sign-change superconducting gap as shown in Figs. \ref{phase} (a) and (b). When $n$ is close to 0.5, the $A_1$ IR with an $s$-wave inter-pocket sign-change superconducting gap is dominant, while for a little smaller $n$, the $d$-wave inter-pocket sign-change gap labeled by $d_{\pm}$ become prominent. Additionally, when $n$ is around 0.2, there exists a region of $A_1$ IR with spin-triplet dominant. In the case where Hubbard $U$ is large shown in Fig. \ref{phase} (c) and (d), spin density wave (SDW) states emerge in the large $n$ region, overshadowing the $A_1$ IR in large $n$ regime, resulting in a transition from spin-triplet-dominant to spin-singlet-dominant types in the $A_1$ gap function, as shown in Figs. \ref{phase} (c) and (d).
	
\begin{figure*}[!htbp]
	\centering
	\includegraphics[width=7in]{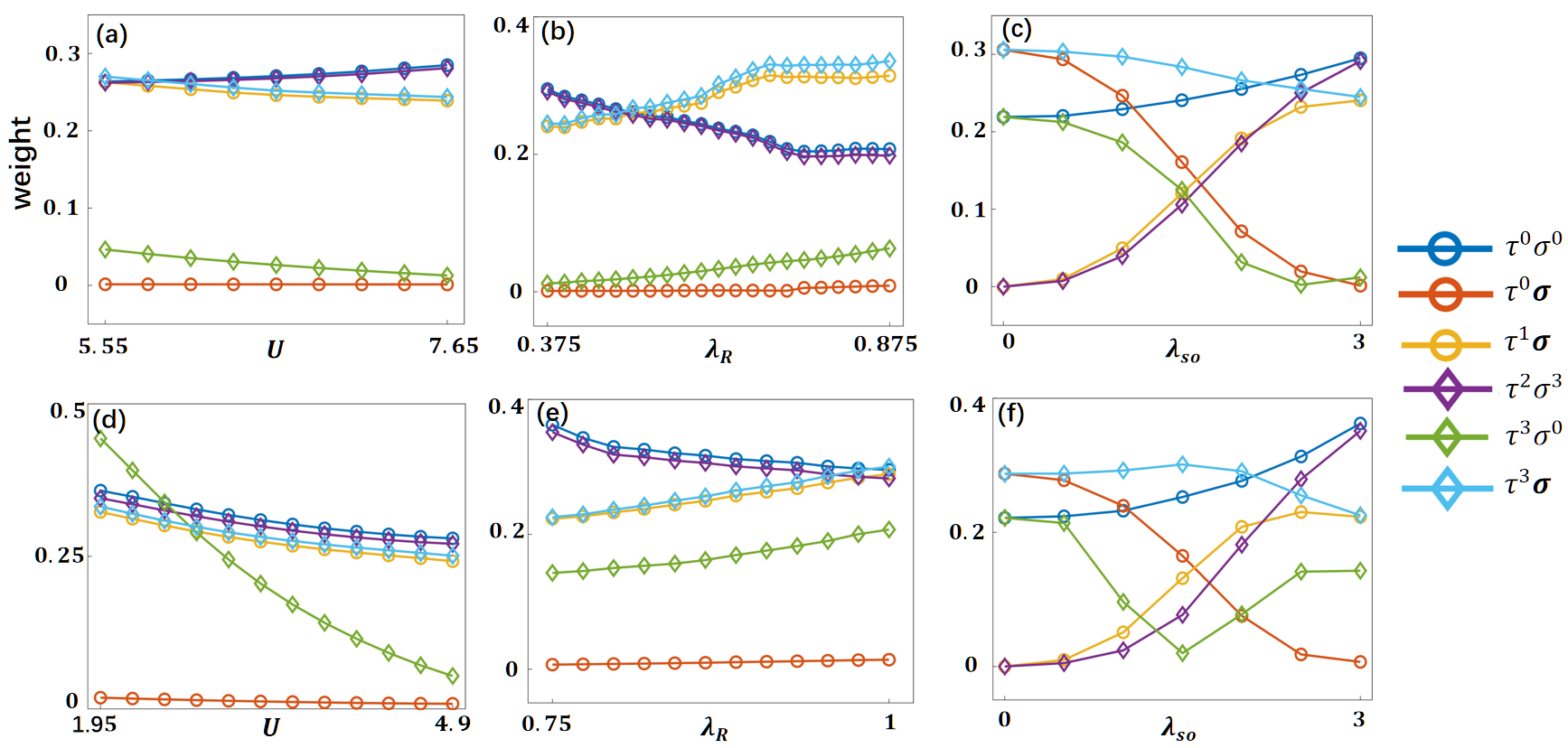}
	\caption{ (a)-(f) The plottings illustrate the contribution weights from six channels to the $A_1 (S_{\pm})$-wave as functions as some specific parameters.  The other fixed parameters are set as flollows: (a) $\lambda_{so}=3$, $n=0.26$ and $\lambda_{R}=0.35$, corresponding to Fig. \ref{fs} (p), (b) $\lambda_{so}=3$, $n=0.26$ and $U=7.5$, (c) $\lambda_{R}=0.35$, $n=0.26$ and $U=7.5$, (d) $\lambda_{so}=3$, $n=0.5$ and $\lambda_{R}=0.9$, (e) $\lambda_{so}=3$, $n=0.5$ and $U=3.5$, (f) $\lambda_{R}=0.725$, $n=0.5$ and $U=3.5$. Note that in (a), (b), (d) and (e), when the condition requiring the absolute value of the weight is not imposed, the pairing in $\tau^3\sigma^0$ channel can acquire a minus sign under a direct projection. In (f), the right part of the green-line minimum is also taken in absolute value for the same reason.  }
	\label{parameter}
\end{figure*}

To investigate the parity-mixed states and elucidate the origin of this $S_{\pm}$-wave pairing, we express the gap function in a parameterized form as follows\cite{gor2001},
\begin{equation}
	\hat{\Delta}_{k}= i\sigma^2[\hat{\phi}_k+\bm{\hat{d}_k}\cdot\bm{\sigma}].
	\label{gap-mix}
\end{equation}
In the one-orbital Rashba model, the projection of the gap function onto the two spin-split Fermi surfaces can be expressed in a simplified form, $\Delta _{1/2}(k)=\phi _{k}\pm |\hat{d}_{k}|$ with $\phi _{k}$ as a scalar. Thus, the aforementioned sign-change gap structure only occur when the triplet component is relatively strong and has helical $p$- or $f$-wave spin-triplet pairing, such as when the filling is near the van Hove singularity\cite{greco2018mechanism, greco2020, bonetti2024}. Any doping away from the van Hove singularity consistently results in predominant nodal d-wave spin-singlet pairing\cite{ nogaki2020}.
However, for the two-orbital model considered here, additional possibilities arise. We project the six sub-channels $i\sigma^2$$\Delta_0$($\tau^0\sigma^0\phi_{1k}$, $\tau^3\sigma^0\phi_{2k}$, $\tau^0\bm{\sigma}\cdot\bm{d_{3k}}$, 
$\tau^3\bm{\sigma}\cdot\bm{d_{4k}}$, $\tau^1\bm{\sigma}\cdot\bm{d_{5k}}$, $\tau^2\sigma^3d_{6k}$) in the spin-orbital representation, listed in Table \ref{pair-class} onto the two Fermi surfaces. The details can be found in Appendix B. Note that the first two sub-channels belong to spin-singlet while the latter four correspond to spin-triplet.

We further distinguish $A_1(S_{\pm})$ states according to the weights of each channel with varying of some parameters, as shown in Fig. \ref{parameter}. This demonstrates that the superconducting state exhibits rich parity-mixed characteristics. Note that the emergence of $B_2(d_{\pm})$ state requires strong RSOC; an example with $\lambda_R=0.95$ is provided in Appendix C.

Figures \ref{parameter} clearly demonstrate that the $S_{\pm}$-wave state exhibits parity-mixed, with the relative contributions of the each channel depending on specific model parameters. This tunability makes our model a promising platform for exploring parity-mixed superconductivity, should it be realized in future material systems. As shown in Figs. \ref{parameter} (a) and (d), the ranking of the contributions from each channel is almost unaffected by Hubbard $U$; however, in certain cases where channel contributions are similar, Hubbard $U$ can alter the leading channel as shown in Fig. \ref{parameter} (a). For $n$=0.26, fig. \ref{parameter} (b) illustrates that when $\lambda_R$ is small, the $S_{\pm}$-wave state is predominantly even-parity channels ($\tau^0\sigma^0$ and $\tau^2\sigma^3$). As $\lambda_R$ increases, their contribution diminishes, while the odd-parity components increase. This suggests that the spin-triplet pairing in this system is primarily induced by $\lambda_R$, fundamentally distinct from the intrinsic $p$-wave or $f$-wave triplet pairing mediated by ferromagnetic fluctuations near a van Hove singularity via doping. In contrast, our $S_{\pm}$-wave state is mediated by antiferromagnetic fluctuations, consistent with this mechanism. For $n$=0.5 as shown in fig. \ref{parameter} (e), the  dominant behavior of even-parity channels becomes even more pronounced, and over a wide range of the parameter $\lambda_R$, the spin-singlet channel remains the leading. This further indicates that the origin of the $S_{\pm}$ state is unrelated to $p$-wave or $f$-wave pairing. Therefore, in both weak and strong RSOC regime, the gap function of the $S_{\pm}$-wave state can be mainly governed by the singlet component, regardless of the presence of triplet contributions—even though the triplet channels can produce the same gap sign distribution, as shown in Fig. \ref{A1p} in Appendix. Figure \ref{parameter} (c) and (f) indicates that $\tau^0\sigma^0$ can be enhanced by $\lambda_{so}$. As discussed in relation to Fig. \ref{fs}, the introduction of $\lambda_{so}$ enhances the contribution of $\tau^0\sigma^0$ by canceling transverse susceptibility, ultimately strengthening inter-pocket interactions, which also enhances another even-parity channel, $\tau^2\sigma^3$.

\section{Topological superconductivity} For noncentrosymmetric systems with time-reversal symmetry, the criteria for identifying TSC require that a pair of spin-split Fermi surfaces enclose an odd number of time-reversal invariant momentum points such as $\Gamma$, $X$ and $M$ points, as shown in Fig. \ref{band} (b), where the Hamiltonian remains invariant under the time-reversal operation, and the order parameter must have opposite signs on the two Fermi surfaces\cite{qi2010}. Additionally, the fully gapped superconducting state is preferable.
According to this criterion,  the Fermi surface in Fig. \ref{fs}  (p), with the $S_{\pm}$-wave pairing we focus on, support TSC.
\begin{figure}[h]
	\centering
	\includegraphics[width=1.0\columnwidth]{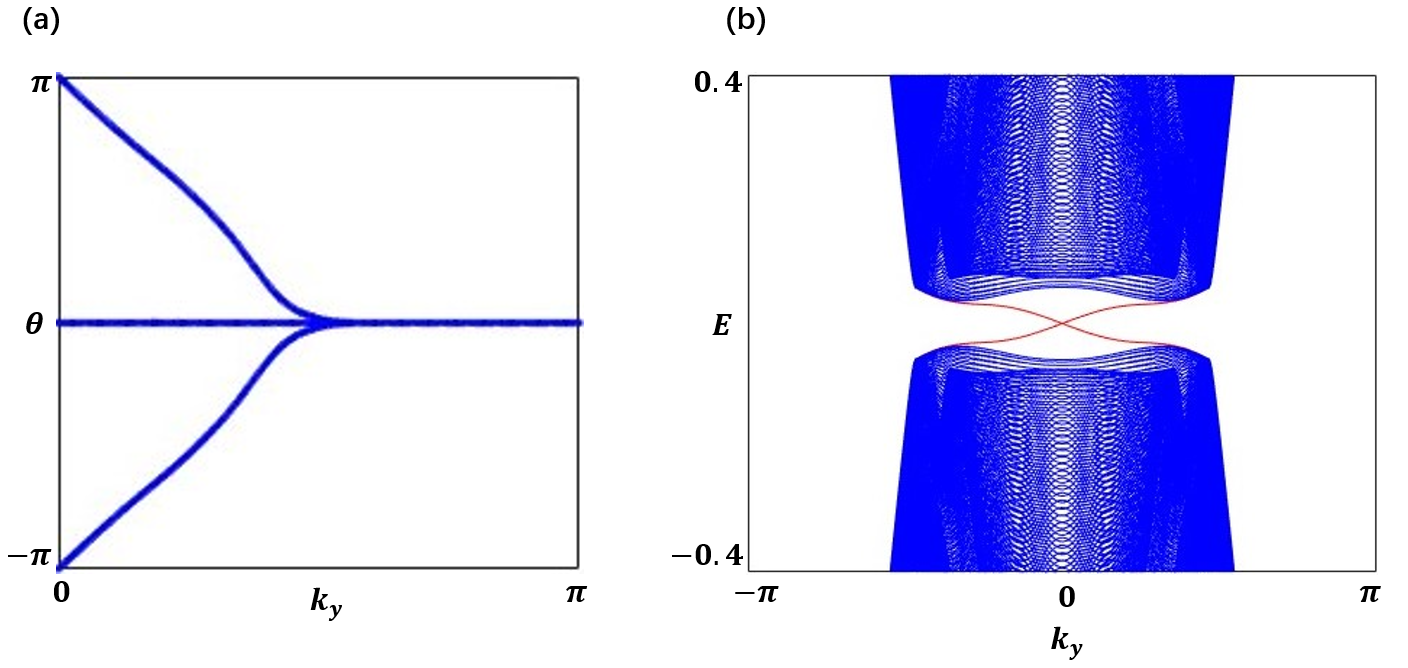}
	\caption{ (a) The Wilson loop\cite{yu2011} calculation for the $S_{\pm}$ superconducting state. (b) The quasi-particle spectrum of $S_{\pm}$ superconducting state.  Note that the in-gap edge states are double-degeneracy with opposite chriality. Periodic and open boundary conditions are applied along $x$ and $y$ direction, respectively. In both (a) and (b), the parameters are the same as those in Fig. \ref{fs} (p). The parameter 
$\Delta_0$ represents the coefficient of the form factors $\phi_{i,k}$ and $\bm{d_{j,k}}$ in the superconducting gap function and is set to 0.1.}  
	\label{z2}
\end{figure}

For the two-dimensional system with time-reversal symmetry considered here, the Chern number is always zero, and the topological nature of the superconducting state can be characterized by a topological invariant $Z_2$. Due to the lack of spatial inversion symmetry, the $Z_2$  invariant can be determined by calculating the Wannier centers of the superconducting state's wave functions\cite{yu2011} and verified through the computation of the edge spectrum\cite{ hao2008, hao2011}. The details can be found in Appendix B. Figs. \ref{z2} (a) and (b) show the Wannier center and edge spectrum, respectively, for the superconducting state with the Fermi surface from Fig. \ref{fs} (p) and the gap structure from Fig. \ref{A1}. It is straightforward to conclude that this superconducting state has a topological invariant $Z_2=1$.  The $B_2(d_{\pm})$ state has nodes, however, it is
also topologically nontrivial\cite{sato20111}, and exhibites Majorana flat bands and dispersive Majorana states on edges (See Appendix C).

\section{Outlook and conclusion.} The TSC state driven by predominant spin-singlet pairing in our study has significant implications. First, TSC states realized through $p$- or $f$-wave spin-triplet pairing are extremely rare in real material systems. In contrast, almost all reported superconductors exhibit spin-singlet pairing, including cuprate, iron-based, and nickelate high-temperature superconductors\cite{bednorz1986, kamihara2008, sun2023}. 
Second, the splitting of Fermi surfaces accompanied by spin-texture characteristics due to broken inversion symmetry has been experimentally reported in systems such as cuprate and iron-based superconductors\cite{gotlieb2018revealing, borisenko2016}. For instance, spin splitting induced by Rashba SOC could potentially be tuned via electric fields or other external controls. Our proposal offers a viable pathway to realize bulk topological superconductivity in high-temperature superconducting systems without relying on proximity effects. If achieved, this could greatly advance the exploration of topological quantum computing based on topological superconductivity.

In summary, we investigated superconductivity in a new class of correlated non-centrosymmetric systems. Using the theory of antiferromagnetic-fluctuation-mediated superconductivity, we identified three possible superconducting states $A_1(S_{\pm})$, $B_2$ and $B_2(d_{\pm})$ under different model parameters. Notably, all states are parity-mixed, and the ratio between even-parity and odd-parity components can be tuned by the model parameters. More importantly, the $A_1(S_{\pm})$ state belongs to a bulk topological superconducting state characterized by a $Z_2=1$ topological invariant.  In both weak and strong RSOC regime, the $A_1(S_{\pm})$  state can be predominantly driven by spin-singlet components, while $B_2(d_{\pm})$ is a nodal TSC and only emerges in strong RSOC regime.  Our proposal is different from all present scenarios based on proximity effects and strong $p$- or $f$-wave spin-triplet pairing. 
\begin{acknowledgments}
	This work was financially supported by the National Key R\&D Program of China (Grants No. 2024YFA1613200, No. 2022YFA1403200 and Grant No. 2023YFA1407300), National Natural Science Foundation of
	China (Grants No. 92265104, No. 12022413 and No. 12447103), the Basic Research Program of the Chinese
	Academy of Sciences Based on Major Scientific Infrastructures (Grant No. JZHKYPT-2021-08), the CASHIPS Director’s Fund (Grant No. BJPY2023A09), Anhui Provincial Major S\&T Project(s202305a12020005), and
	the High Magnetic Field Laboratory of Anhui Province under Contract No. AHHM-FX-2020-02. 
\end{acknowledgments}
\renewcommand{\thefigure}{A\arabic{figure}}
\setcounter{figure}{0}
\appendix
		\section{Formalism of the multi-orbital RPA approach \label{RPA}}
		\begin{widetext}
		We calculate the susceptibility using the expression
		\begin{equation}
			\chi^{l_1\sigma_1l_2\sigma_2}_{l_3\sigma_3l_4\sigma_4}(\bm{q},i\nu_n)=\sum_{\bm{k},\bm{k}'}\int_{0}^{\beta}d\tau\braket{T_{\tau}
				c^{\dagger}_{l_1\sigma_1}(\bm{k}+\bm{q},\tau)c_{l_2\sigma_2}(\bm{k},)c^{\dagger}_{l_4\sigma_4}(\bm{k}'-\bm{q},0)c_{l_3\sigma_3}(\bm{k}',0)}
			\nonumber
		\end{equation}
		The Green's function expression of the lowest order is given by
		\begin{equation}
			\chi^{l_1\sigma_1l_2\sigma_2}_{l_3\sigma_3l_4\sigma_4}(\bm{q},i\nu_n)
			=-T\sum_{\bm{k},i\omega_n}G_{l_3\sigma_3l_1\sigma_1}(\bm{k},i\omega_n)G_{l_2\sigma_2l_4\sigma_4}(\bm{k}+\bm{q},i\omega_n+i\nu_n),
			\label{nonumber}
		\end{equation}
	\begin{equation}
		\chi^{l_1\sigma_1l_2\sigma_2}_{l_3\sigma_3l_4\sigma_4}(\bm{q},i\nu_n)=-\frac{1}{N}\sum_{\bm{k},\mu\nu}a^{l_3\sigma_3}_{\mu}(\bm{k})a^{l_1\sigma_1*}_{\mu}(\bm{k})a^{l_2\sigma_2}_{\nu}(\bm{k}+\bm{q})a^{l_4\sigma_4*}_{\nu}(\bm{k}+\bm{q})
		\frac{f(E_{\nu}(\bm{k}+\bm{q}))-f(E_{\mu}(\bm{k}))}{E_{\nu}(\bm{k}+\bm{q})-E_{\mu}(\bm{k})+i\nu_n}
		\label{sus}
	\end{equation}
		\end{widetext}
		where the Matsubara Green's function is
		\begin{equation}
			\begin{aligned}
				\hat{G}(\bm{k},i\omega_n)=[i\omega_n-H_0(\bm{k})]^{-1}.
			\end{aligned}
			\label{greenfunc}
			\nonumber
		\end{equation}
		Here, the indices $l_{1-4}$ refer to orbital indices, while the indices $\sigma_{1-4}$ refer to spin indices. $i\omega_{n}$ and $i\nu_n$ denote Fermionic and Bosonic Matsubara frequencies, respectively.
		In practical calculations, we can express the susceptibility as Eq. (\ref{sus})
		where $N$ is the number of $\bm{k}$-points in the Brillouin Zone, $a^{l\sigma}_{\mu}(\bm{k})=\braket{l\sigma,\bm{k}|\mu,\bm{k}}$ is the wave function in energy space, and $f(E)$ is the Fermi-Dirac distribution. For the calculation of static susceptibility, we set $i\nu_n=i0^+$ on the right side of equation (\ref{sus}), while on the left side, we have $\chi^{l_1\sigma_1l_2\sigma_2}_{l_3\sigma_3l_4\sigma_4}(\bm{q},i\nu_n=0)$ to smooth results of the calculation.
		
		The Hubbard interaction matrix is given by
		\begin{equation}
			\begin{aligned}
				&W^{l_1\sigma l_1\bar{\sigma}}_{l_1\sigma l_1\bar{\sigma}}=U && W^{l_1\sigma l_1\sigma}_{l_1\bar{\sigma} l_1\bar{\sigma}}=-U\\
				&W^{l_2\sigma l_1\bar{\sigma}}_{l_2\sigma l_1\bar{\sigma}}=V && W^{l_1\sigma l_1\sigma}_{l_2\bar{\sigma} l_2\bar{\sigma}}=-V\\
				&W^{l_1\sigma l_1\bar{\sigma}}_{l_2\sigma l_2\bar{\sigma}}=J && W^{l_1\sigma l_2\sigma}_{l_1\bar{\sigma} l_2\bar{\sigma}}=-J\\
				&W^{l_1\sigma l_2\bar{\sigma}}_{l_1\sigma l_2\bar{\sigma}}=J' && W^{l_1\sigma l_2\sigma}_{l_2\bar{\sigma} l_1\bar{\sigma}}=-J'\\
			\end{aligned}
			\nonumber
		\end{equation}
		with all other matrix elements equal to zero. The RPA susceptibility is given by
		$$
		\chi^{RPA}(\bm{q})=[1-\chi(\bm{q})W]^{-1}\chi(\bm{q}),
		$$
		where the charge susceptibility is
		$$
		\chi^{RPA}_c(\bm{q})=\sum_{l_1,l_2}\sum_{\{\sigma\}}\sigma^0_{\sigma_1\sigma_2}\sigma^0_{\sigma_4\sigma_3}\chi^{l_1\sigma_1l_1\sigma_2}_{l_2\sigma_3l_2\sigma_4}(\bm{q}),
		$$
		the longitudinal susceptibility is
		$$
		\chi^{RPA}_{zz}(\bm{q})=\sum_{l_1,l_2}\sum_{\{\sigma\}}\sigma^z_{\sigma_1\sigma_2}\sigma^z_{\sigma_4\sigma_3}\chi^{l_1\sigma_1l_1\sigma_2}_{l_2\sigma_3l_2\sigma_4}(\bm{q}),
		$$
		and the transverse susceptibility is
		$$
		\chi^{RPA}_{+-}(\bm{q})=\sum_{l_1,l_2}\sum_{\{\sigma\}}\sigma^+_{\sigma_1\sigma_2}\sigma^{-}_{\sigma_4\sigma_3}\chi^{l_1\sigma_1l_1\sigma_2}_{l_2\sigma_3l_2\sigma_4}(\bm{q}),
		$$
		
\begin{figure}[h]
			\centering
			\includegraphics[width=3.5in]{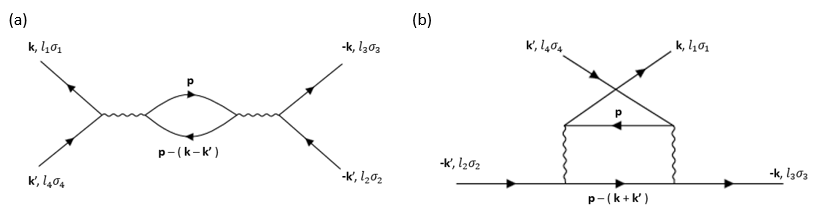}
			\caption{The second-order bubble and ladder diagram}
			\label{Feynman}
		\end{figure}

		After applying the RPA, we are able to calculate the effective interaction of Cooper pairs in orbital-spin space. The relevant Feynman diagrams include ladder and bubble diagrams, which are illustrated in FIG. \ref{Feynman}. The total effective pairing interaction can be expressed as
		\begin{equation}
			V^{l_1\sigma_1l_2\sigma_2}_{l_3\sigma_3l_4\sigma_4}(\bm{k},\bm{k'})=W^{l_1\sigma_1l_2\sigma_2}_{l_3\sigma_3l_4\sigma_4}+V_{l}(\bm{k},\bm{k'})-V_{b}(\bm{k},\bm{k'}),
			\nonumber
		\end{equation}
		with the ladder term in FIG. \ref{Feynman} (b)
		\begin{equation}
			V_{l}(\bm{k},\bm{k'})=[\hat{W}\hat{\chi}^{RPA}\hat{W}]^{l_1\sigma_1l_2\sigma_2}_{l_3\sigma_3l_4\sigma_4}(\bm{k}+\bm{k'}),
			\nonumber
		\end{equation}
		and the bubble term in FIG. \ref{Feynman} (a)
		\begin{equation}
			V_{b}(\bm{k},\bm{k'})=[\hat{W}\hat{\chi}^{RPA}\hat{W}]^{l_1\sigma_1l_4\sigma_4}_{l_3\sigma_3l_2\sigma_2}(\bm{k}-\bm{k'}).
			\nonumber
		\end{equation}
		The potential superconducting pairing can be evaluated by solving the following linearized Eliashberg equation,
		\begin{widetext}
		\begin{equation}
			\lambda\Delta_{l_1\sigma_1l_4\sigma_4}(\bm{k})
			=T\sum_{\bm{k'}i\omega_n}\sum_{l_2l_3\sigma_2\sigma_3}V^{l_1\sigma_1l_2\sigma_2}_{l_3\sigma_3l_4\sigma_4}(\bm{k},\bm{k'})F_{l_3\sigma_3l_2\sigma_2}(\bm{k'},i\omega_n) 
			\label{gf}
		\end{equation}
		with
		\begin{equation}
			F_{l_3\sigma_3l_2\sigma_2}(\bm{k'},i\omega_n)
			=G_{l_3\sigma_3l\sigma}(\bm{k'},i\omega_n)\Delta_{l\sigma l'\sigma'}(\bm{k'})G_{l_2\sigma_2l'\sigma'}(-\bm{k'},-i\omega_n).
			\nonumber
		\end{equation}
	\end{widetext}
		Here, $\Delta_{l_1\sigma_1l_4\sigma_4}(\bm{k})$ is the superconducting pairing function, and the leading superconducting instability can be identified by the largest eigenvalue $\lambda_{\alpha}$.
		\section{BdG Hamiltonian projection between spin-orbital space and band space, and calculation of $Z_2$ \label{Z2}}
		We now focus on the $S_{\pm}$-wave states in orbital-spin space. 
		The BdG Hamiltonian is given by
		\begin{equation}
			H_{BdG}= \left[
			\begin{aligned}
				H_0(k) && \Delta(k) \\
				\Delta^{*}(k) && -H^*_0(-k)
			\end{aligned}  
			\right].  
			\nonumber
		\end{equation}
		
		Here, the normal part $H_0(k)$ can be can be diagonalized using $\hat{V}^{-1}(k)H_0(k)\hat{V}(k)=\hat{E}(k)$, with $\hat{E}_{\mu\nu}=E_{\mu}(k)\delta_{\mu\nu}$. $\mu$ and $\nu$ are the band indices. $\hat{V}(k)$  can be parameterized as follows:
		\begin{equation}
			\hat{V}(k)=
			\left[
			\begin{aligned}
				C_1e^{i\theta} && -C_3e^{i\theta} && C_4e^{i\theta} && C_2e^{i\theta}\\
				-iC_1 && -iC_3 && iC_4 && iC_2 \\
				-iC_2e^{i\theta} && iC_4e^{i\theta} && iC_3e^{i\theta} && -iC_1e^{i\theta}\\
				C_2 && C_4 && C_3 && C_1
			\end{aligned}
			\right],
			\nonumber
		\end{equation}
		where $e^{i\theta}=(\sin k_y+i\sin k_x)/h_{\bm{k}}$, $2(C_1^2+C_2^2)=2(C_3^2+C_4^2)=1$ and $h_{\bm{k}}=\sqrt{\sin^2k_x+\sin^2k_y}$.
		
\begin{figure*}[!htbp]
			\centering
			\includegraphics[width=7in]{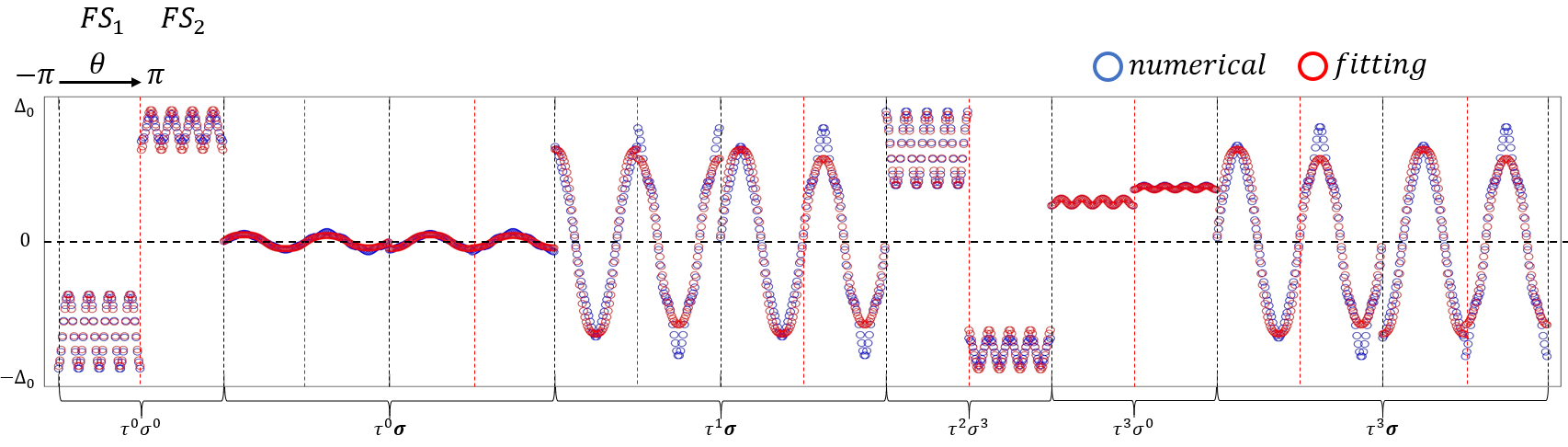}
			\caption{Gap function in six possible channel of $A_1$ IR under the condition of Fig. \ref{fs} (p) over the angle $\theta$ on the two Fermi surfaces.  Here, the defination of polar angle $\theta$ can be found in Fig. \ref{A1p} (a). $\Delta_0$ is a constant determined by a cut-off energy.}
			\label{A1}
		\end{figure*}

		For the pairing function belonging to the $A_1$ irreducible representation, it can be rewritten as
		\begin{widetext}
		\begin{equation}
			\begin{aligned}
				\Delta(k)/(i\sigma^2\Delta_0)&=\tau^0\sigma^0\phi_{1k}+\tau^0\bm{\sigma}\cdot \bm{d_{3k}}
				+\tau^2\sigma^3d_{6k}+\tau^1\bm{\sigma}\cdot \bm{d_{5k}}
				+\tau^3\sigma^0\phi_{2k}+\tau^3\bm{\sigma}\cdot \bm{d_{4k}}\\
			\end{aligned} 
			\label{so1}
		\end{equation}
		\begin{equation}
		\begin{aligned}
			&\phi_{1k}=-0.179+0.093(\cos k_{x}+\cos k_{y})+0.189\cos k_{x}\cos k_{y}&\\
			&\bm{d_{3k}}=-0.003(\sin k_y, -\sin k_x, 0)&\\
			&\bm{d_{5k}}=-0.04(\sin k_x, \sin k_y, 0)&\\
			&d_{6k}=(0, 0, 0.182-0.095(\cos k_{x}+\cos k_{y})-0.186\cos k_{x}\cos k_{y})&\\
			&\phi_{2k}=0.011+0.002(\cos k_{x}+\cos k_{y})+0.019\cos k_{x}\cos k_{y}&\\	
			&\bm{d_{4k}}=-0.04(\sin k_y, -\sin k_x, 0)&\\
		\end{aligned}
		\label{ch}
	\end{equation}
	\end{widetext}
		Here, $\phi_{1k}$, $\phi_{2k}$ and $d_{6k}$ has general form of extended s-wave gap function, $\bm{d_{3k}}= \bm{d_{4k}}\propto (-sink_y,sink_x,0)$ and $\bm{d_{5k}}\propto (sink_x,sink_y,0)$. By solving the Eliashberg function in Eq. (\ref{gf}), the numerical results in orbital-spin space can be obtained and are presented in Fig. \ref{A1}. The fitting results as listed as Eq. (\ref{ch})
		\begin{figure*}[!htbp]
			\centering
			\includegraphics[width=5.5in]{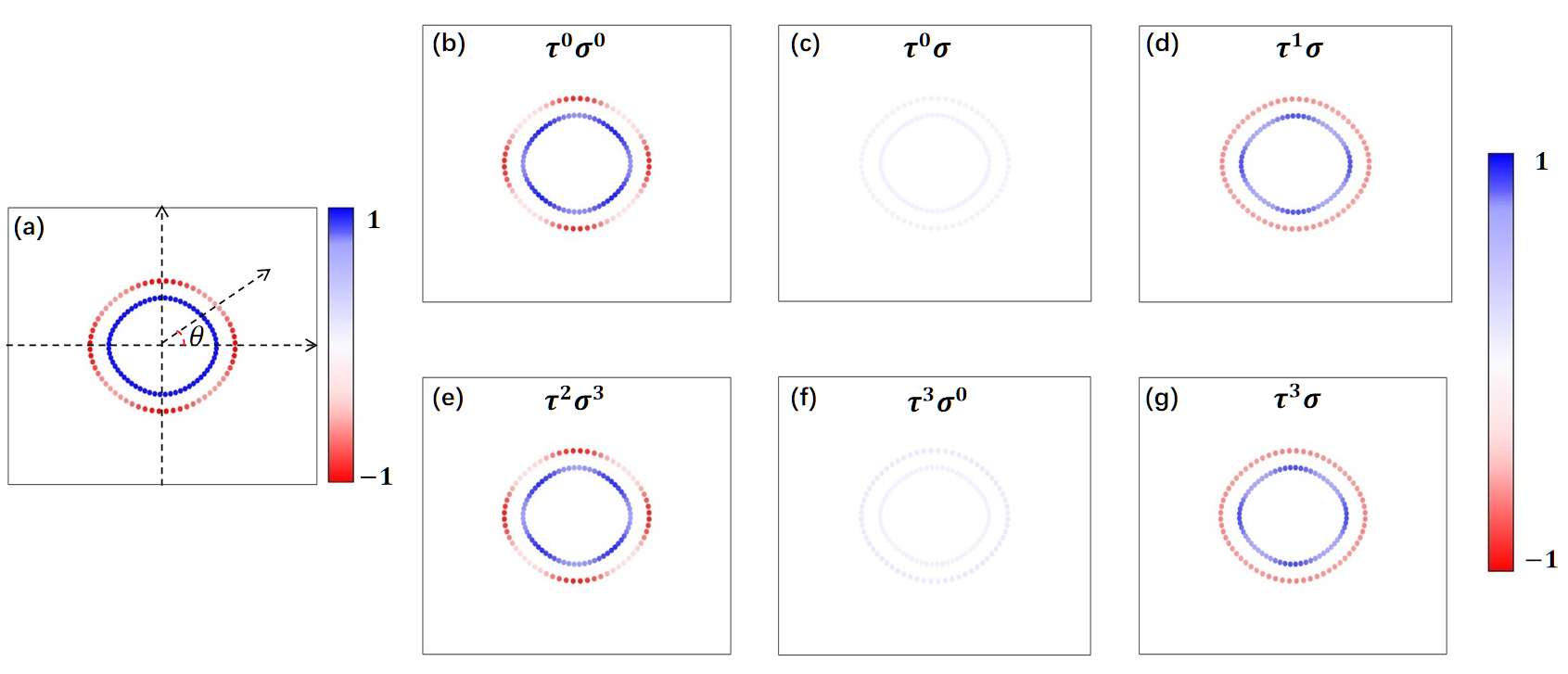}
			\caption{ (a) The projected gap function of full $A_1$ IR in band space, normalized to its maximum. Here, the polar angle $\theta$ is defined in the polar coordinate system.  (b)-(g) The projected gap function of six  individual  sub-channels of $A_1$ IR in band space, renormalized by the maximum of $i\sigma^2\tau^0\sigma^0\phi_{1k}$. Here U=7.5 and J=0. Other parameters are the same with Fig. \ref{fs} (p) in main text. }
			\label{A1p}
		\end{figure*}
			
		Do the projection, the effective gap function in band space of $H_0$ is given by
		\begin{equation}
			\Delta_{1/2}(k)=\Delta_0(\phi_{1k}+2(C_{1/3}^2-C_{2/4}^2)\phi_{2k}-4C_{1/3}C_{2/4}d_{6k}\mp C^{t}_{1/2})
			\nonumber
		\end{equation}
		where
		\begin{equation}
			\begin{aligned}
				&C^{t}_{1/2}=2(C_{1/3}^2-C_{2/4}^2)|d_{3k}|-4C_{1/3}C_{2/4}|d_{5k}|+|d_{4k}|
			\end{aligned}
			\nonumber
		\end{equation}
		in this case, $C_2/C_1<<1$, $C_4/C_3<<1$, then we have
		\begin{equation}
			\Delta_{1/2}(k)\approx \Delta_0(\phi_{1k}+\phi_{2k}-2C_{2/4}d_{6k}\mp C^{t}_{1/2}) 
			\label{b1}
		\end{equation}
		where
		\begin{equation}
			\begin{aligned}
				&C^{t}_{1/2}\approx |d_{3k}|-2C_{2/4}|d_{5k}|+|d_{4k}|
			\end{aligned}
			\nonumber
		\end{equation}
		
Note that the gap functions in Eq. \ref{so1} and Eq. \ref{b1} are expressed in spin-orbital and band basis, respectively. They exhibit a one-to-one correspondence. In Fig. \ref{A1p} (a), we present the distribution of the total projected gap function on the Fermi surface in the band basis, showing a clear $S_{\pm}$-wave characteristic. In Figs. \ref{A1p} b–g, we show the distributions of the projected gap function on the Fermi surface in the band basis when each subchannel in the spin-orbital basis is individually nonzero. It can be seen that the two spin-singlet subchannels carry the strongest weight, and all subchannels share the same $S_{\pm}$-wave characteristic.  Note that for the remaining two bands, which lie far from the Fermi level, the gap function acquires an opposite sign; however, these contributions can be safely ignored since superconductivity is governed primarily by the low-energy states near the Fermi surface.
		
Given that the system is time-reversal invariant, the Chern number is always zero. Therefore we have to use the $Z_2$ number to study its topological properties. We employ Wilson's loop approach \cite{PhysRevB.84.075119} where the non-Abelian Berry connection is defined as
		
	\begin{equation}
			F^{mn}_{i,i+1}=\braket{n,k_{x,i},k_y|m,k_{x,i+1},k_y}.
			\label{Berry1}
		\end{equation}
		
	Multiplying $F^{mn}_{i,i+1}$ from A to Z including $F^{mn}_{N,1}$, a new matirx $D(k_y)$ can be obtained as follow, 
\begin{equation}
		D(k_y)=F_{1,2}F_{2,3}...F_{N,1}.
		\label{Berry3}
\end{equation}
Here, $D(k_y)$ is a $4 \times 4$ matrix whose eigenvalue phases $\theta^{D}_{m}(k_y)$ represent the evolution of the Wannier center pair in $k_y \in [-\pi, \pi]$. The integer winding number of the Wannier center pair can be viewed as the number of times it encloses the original point.
			\begin{figure*}[!htbp]
			\centering
			\includegraphics[width=6in]{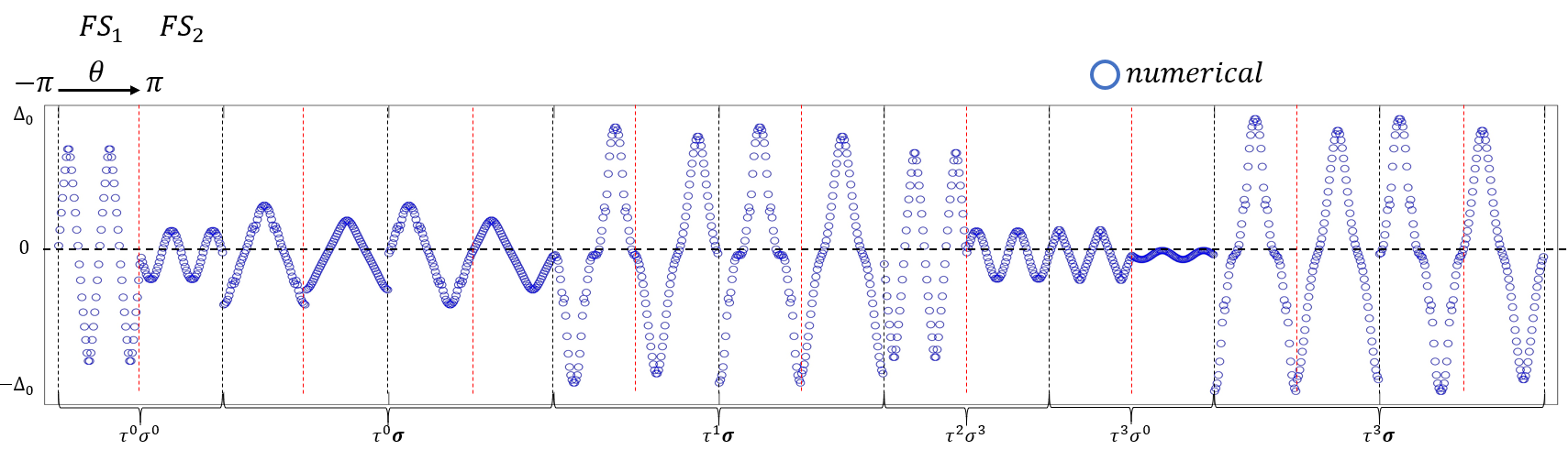}
			\caption{The distribution of the gap functions for the six subchannels in the $d_{\pm}$-wave states as a function of the angle $\theta$ on the two Fermi surfaces is presented. The parameters used are $\lambda_R=0.95$, $\lambda_{so}=3$,  $n=0.26$, $U=3.5$ and other parameters remain consistent with main text.}
			\label{dpmfit}
		\end{figure*}
		
		\begin{figure*}[!htbp]
			\centering
			\includegraphics[width=6in]{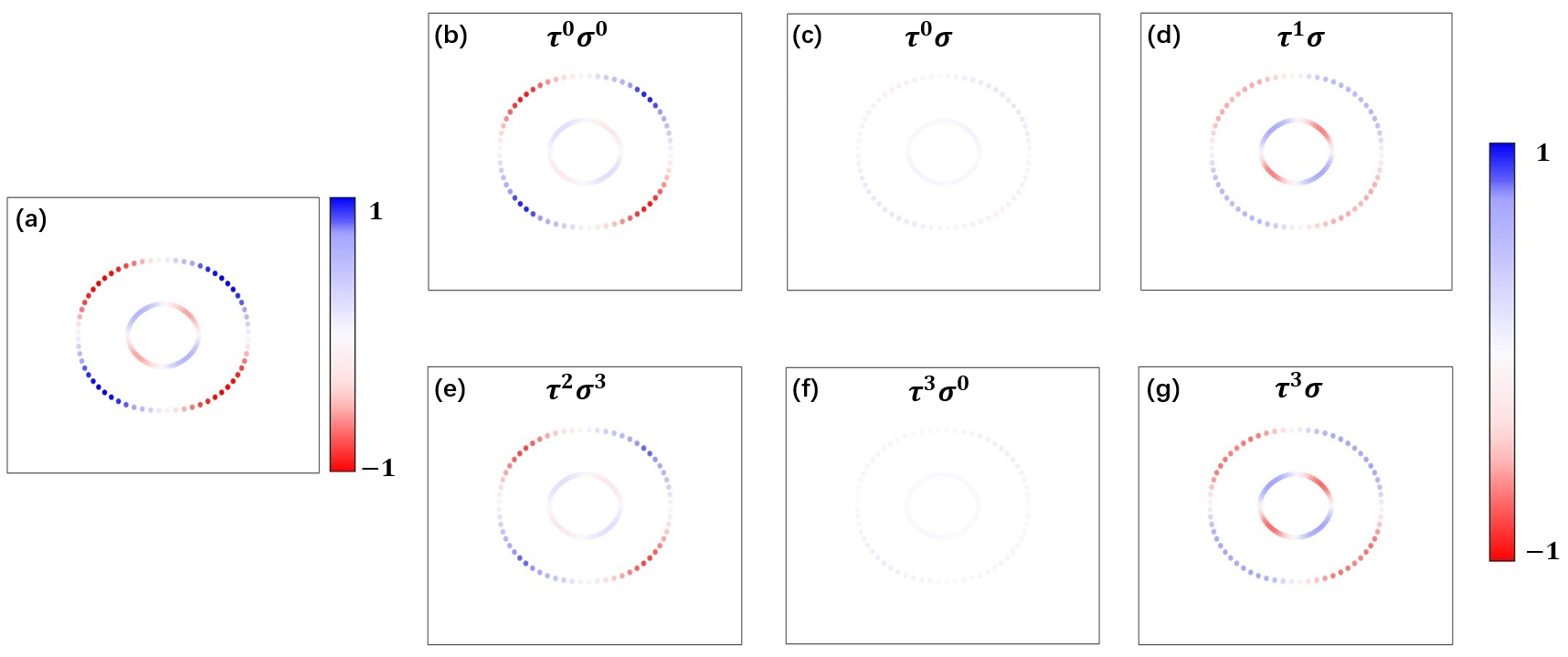}
			\caption{(a) The projected gap function of full $B_2$ IR in band space, normalized to its maximum. (b)-(g) The projected gap function of six  individual  sub-channels of $B_2$ IR in band space, renormalized by the maximum of $i\sigma^2\tau^0\sigma^0\phi_{1k}$. Here parameters are the same with Fig. \ref{dpmfit}.}
			\label{dpmchannel}
		\end{figure*}
		\begin{figure}[!]
			\centering
			\includegraphics[width=3in]{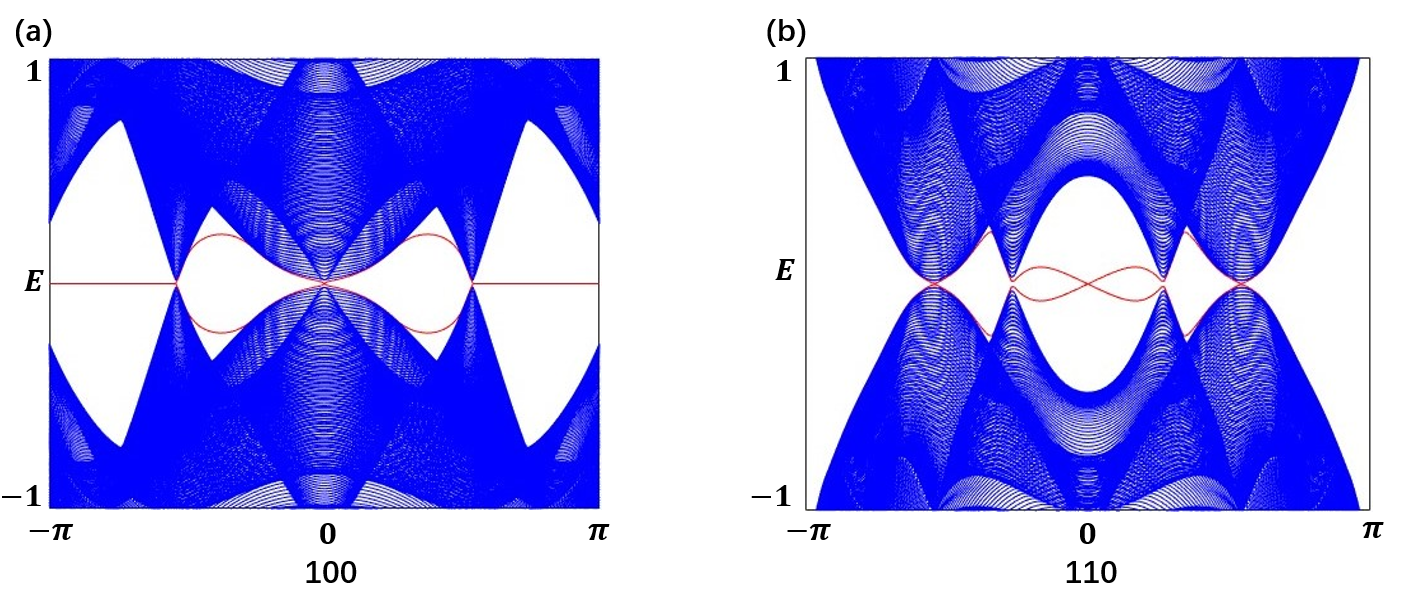}
			\caption{(a) The quasi-particle spectrum of $d_{\pm}$ superconducting state in (100) direction. (b) The quasi-particle spectrum of $d_{\pm}$ superconducting state in (110) direction.}
			\label{dpmedge}
		\end{figure}
		\section{Properties of $d_{\pm}$ waves \label{dpm}}
		An addtional noteworthy class of superconducting gaps identified in our calculations is the $B_{2}(d_{\pm})$ wave. To obtain this result, we increased the strength of the RSOC to $\lambda_R=0.95$, surpassing the onsite SOC of $\lambda_{so}=3$. Figure (\ref{dpmfit}) clearly demonstrates that the wave state exhibits parity mixing, indicating that the sign changes across different Fermi surfaces are not solely attributed to spin-triplet subchannels; rather, spin-singlet subchannels inherently exhibit sign changes as well. In the spin-orbital representation, the superconducting gap function can be fitted up to 6th-nearest-neighbor terms. Specifically, $\phi_{1k}=0.138 \sin k_x \sin k_y-0.005\sin 2k_x \sin 2k_y - 0.05 (\sin 2k_x \sin k_y + \sin k_x \sin 2k_y)$.

In Fig. \ref{dpmchannel} (a), we present the distribution of the total projected gap function on the Fermi surface in the band basis, showing a clear $d_{\pm}$-wave characteristic. In Figs. \ref{dpmchannel} b–g, we show the distributions of the projected gap function on the Fermi surface in the band basis. It can be seen that the two spin-singlet subchannels is naturely sign changed and carry strong weight, and all subchannels share the same $d_{\pm}$-wave characteristic.
		
		Regarding the spectrum for a boundary that breaks translational periodicity along the (100)
			direction, two types of edge states appear. The first is a conventional flat-band edge state, analogous to that found in standard $B_2$ pairing. The second is a pair of Dirac-type edge states, which arise from the sign change of the superconducting gap between different Fermi surfaces, as shown in Fig. \ref{dpmedge}(a). It is important to note that these Dirac-type edge states may not always be directly observable, as they can be obscured by bulk states. However, when the boundary is oriented along an alternative direction, such as (110), these edge states can always be detected, since there is no node at $k_{(110)} = 0$ as demonstrated in Fig. \ref{dpmedge}(b).

\begin{thebibliography}{54}%
\makeatletter
\providecommand \@ifxundefined [1]{%
 \@ifx{#1\undefined}
}%
\providecommand \@ifnum [1]{%
 \ifnum #1\expandafter \@firstoftwo
 \else \expandafter \@secondoftwo
 \fi
}%
\providecommand \@ifx [1]{%
 \ifx #1\expandafter \@firstoftwo
 \else \expandafter \@secondoftwo
 \fi
}%
\providecommand \natexlab [1]{#1}%
\providecommand \enquote  [1]{``#1''}%
\providecommand \bibnamefont  [1]{#1}%
\providecommand \bibfnamefont [1]{#1}%
\providecommand \citenamefont [1]{#1}%
\providecommand \href@noop [0]{\@secondoftwo}%
\providecommand \href [0]{\begingroup \@sanitize@url \@href}%
\providecommand \@href[1]{\@@startlink{#1}\@@href}%
\providecommand \@@href[1]{\endgroup#1\@@endlink}%
\providecommand \@sanitize@url [0]{\catcode `\\12\catcode `\$12\catcode
  `\&12\catcode `\#12\catcode `\^12\catcode `\_12\catcode `\%12\relax}%
\providecommand \@@startlink[1]{}%
\providecommand \@@endlink[0]{}%
\providecommand \url  [0]{\begingroup\@sanitize@url \@url }%
\providecommand \@url [1]{\endgroup\@href {#1}{\urlprefix }}%
\providecommand \urlprefix  [0]{URL }%
\providecommand \Eprint [0]{\href }%
\providecommand \doibase [0]{https://doi.org/}%
\providecommand \selectlanguage [0]{\@gobble}%
\providecommand \bibinfo  [0]{\@secondoftwo}%
\providecommand \bibfield  [0]{\@secondoftwo}%
\providecommand \translation [1]{[#1]}%
\providecommand \BibitemOpen [0]{}%
\providecommand \bibitemStop [0]{}%
\providecommand \bibitemNoStop [0]{.\EOS\space}%
\providecommand \EOS [0]{\spacefactor3000\relax}%
\providecommand \BibitemShut  [1]{\csname bibitem#1\endcsname}%
\let\auto@bib@innerbib\@empty
\bibitem [{\citenamefont {Ivanov}(2001)}]{ivanov2001non}%
  \BibitemOpen
  \bibfield  {author} {\bibinfo {author} {\bibfnamefont {D.~A.}\ \bibnamefont
  {Ivanov}},\ }\bibfield  {title} {\bibinfo {title} {Non-abelian statistics of
  half-quantum vortices in p-wave superconductors},\ }\href@noop {} {\bibfield
  {journal} {\bibinfo  {journal} {Physical review letters}\ }\textbf {\bibinfo
  {volume} {86}},\ \bibinfo {pages} {268} (\bibinfo {year} {2001})}\BibitemShut
  {NoStop}%
\bibitem [{\citenamefont {Kitaev}(2003)}]{kitaev20031}%
  \BibitemOpen
  \bibfield  {author} {\bibinfo {author} {\bibfnamefont {A.~Y.}\ \bibnamefont
  {Kitaev}},\ }\bibfield  {title} {\bibinfo {title} {Fault-tolerant quantum
  computation by anyons},\ }\href@noop {} {\bibfield  {journal} {\bibinfo
  {journal} {Annals of physics}\ }\textbf {\bibinfo {volume} {303}},\ \bibinfo
  {pages} {2} (\bibinfo {year} {2003})}\BibitemShut {NoStop}%
\bibitem [{\citenamefont {Kitaev}(2006)}]{kitaev2006anyons}%
  \BibitemOpen
  \bibfield  {author} {\bibinfo {author} {\bibfnamefont {A.}~\bibnamefont
  {Kitaev}},\ }\bibfield  {title} {\bibinfo {title} {Anyons in an exactly
  solved model and beyond},\ }\href@noop {} {\bibfield  {journal} {\bibinfo
  {journal} {Annals of Physics}\ }\textbf {\bibinfo {volume} {321}},\ \bibinfo
  {pages} {2} (\bibinfo {year} {2006})}\BibitemShut {NoStop}%
\bibitem [{\citenamefont {Nayak}\ \emph {et~al.}(2008)\citenamefont {Nayak},
  \citenamefont {Simon}, \citenamefont {Stern}, \citenamefont {Freedman},\ and\
  \citenamefont {Das~Sarma}}]{nayak2008non}%
  \BibitemOpen
  \bibfield  {author} {\bibinfo {author} {\bibfnamefont {C.}~\bibnamefont
  {Nayak}}, \bibinfo {author} {\bibfnamefont {S.~H.}\ \bibnamefont {Simon}},
  \bibinfo {author} {\bibfnamefont {A.}~\bibnamefont {Stern}}, \bibinfo
  {author} {\bibfnamefont {M.}~\bibnamefont {Freedman}},\ and\ \bibinfo
  {author} {\bibfnamefont {S.}~\bibnamefont {Das~Sarma}},\ }\bibfield  {title}
  {\bibinfo {title} {Non-abelian anyons and topological quantum computation},\
  }\href@noop {} {\bibfield  {journal} {\bibinfo  {journal} {Reviews of Modern
  Physics}\ }\textbf {\bibinfo {volume} {80}},\ \bibinfo {pages} {1083}
  (\bibinfo {year} {2008})}\BibitemShut {NoStop}%
\bibitem [{\citenamefont {Alicea}(2012)}]{alicea2012new}%
  \BibitemOpen
  \bibfield  {author} {\bibinfo {author} {\bibfnamefont {J.}~\bibnamefont
  {Alicea}},\ }\bibfield  {title} {\bibinfo {title} {New directions in the
  pursuit of majorana fermions in solid state systems},\ }\href@noop {}
  {\bibfield  {journal} {\bibinfo  {journal} {Reports on progress in physics}\
  }\textbf {\bibinfo {volume} {75}},\ \bibinfo {pages} {076501} (\bibinfo
  {year} {2012})}\BibitemShut {NoStop}%
\bibitem [{\citenamefont {Elliott}\ and\ \citenamefont
  {Franz}(2015)}]{elliott2015colloquium}%
  \BibitemOpen
  \bibfield  {author} {\bibinfo {author} {\bibfnamefont {S.~R.}\ \bibnamefont
  {Elliott}}\ and\ \bibinfo {author} {\bibfnamefont {M.}~\bibnamefont
  {Franz}},\ }\bibfield  {title} {\bibinfo {title} {Colloquium: Majorana
  fermions in nuclear, particle, and solid-state physics},\ }\href@noop {}
  {\bibfield  {journal} {\bibinfo  {journal} {Reviews of Modern Physics}\
  }\textbf {\bibinfo {volume} {87}},\ \bibinfo {pages} {137} (\bibinfo {year}
  {2015})}\BibitemShut {NoStop}%
\bibitem [{\citenamefont {Hasan}\ and\ \citenamefont
  {Kane}(2010)}]{RevModPhys.82.3045}%
  \BibitemOpen
  \bibfield  {author} {\bibinfo {author} {\bibfnamefont {M.~Z.}\ \bibnamefont
  {Hasan}}\ and\ \bibinfo {author} {\bibfnamefont {C.~L.}\ \bibnamefont
  {Kane}},\ }\bibfield  {title} {\bibinfo {title} {Colloquium: Topological
  insulators},\ }\href@noop {} {\bibfield  {journal} {\bibinfo  {journal} {Rev.
  Mod. Phys.}\ }\textbf {\bibinfo {volume} {82}},\ \bibinfo {pages} {3045}
  (\bibinfo {year} {2010})}\BibitemShut {NoStop}%
\bibitem [{\citenamefont {Qi}\ and\ \citenamefont
  {Zhang}(2011)}]{RevModPhys.83.1057}%
  \BibitemOpen
  \bibfield  {author} {\bibinfo {author} {\bibfnamefont {X.-L.}\ \bibnamefont
  {Qi}}\ and\ \bibinfo {author} {\bibfnamefont {S.-C.}\ \bibnamefont {Zhang}},\
  }\bibfield  {title} {\bibinfo {title} {Topological insulators and
  superconductors},\ }\href@noop {} {\bibfield  {journal} {\bibinfo  {journal}
  {Rev. Mod. Phys.}\ }\textbf {\bibinfo {volume} {83}},\ \bibinfo {pages}
  {1057} (\bibinfo {year} {2011})}\BibitemShut {NoStop}%
\bibitem [{\citenamefont {Read}\ and\ \citenamefont {Green}(2000)}]{PRB-1}%
  \BibitemOpen
  \bibfield  {author} {\bibinfo {author} {\bibfnamefont {N.}~\bibnamefont
  {Read}}\ and\ \bibinfo {author} {\bibfnamefont {D.}~\bibnamefont {Green}},\
  }\bibfield  {title} {\bibinfo {title} {Paired states of fermions in two
  dimensions with breaking of parity and time-reversal symmetries and the
  fractional quantum hall effect},\ }\href@noop {} {\bibfield  {journal}
  {\bibinfo  {journal} {Phys. Rev. B}\ }\textbf {\bibinfo {volume} {61}},\
  \bibinfo {pages} {10267} (\bibinfo {year} {2000})}\BibitemShut {NoStop}%
\bibitem [{\citenamefont {Fu}\ and\ \citenamefont
  {Kane}(2008)}]{fu2008superconducting}%
  \BibitemOpen
  \bibfield  {author} {\bibinfo {author} {\bibfnamefont {L.}~\bibnamefont
  {Fu}}\ and\ \bibinfo {author} {\bibfnamefont {C.~L.}\ \bibnamefont {Kane}},\
  }\bibfield  {title} {\bibinfo {title} {Superconducting proximity effect and
  majorana fermions at the surface of a topological insulator},\ }\href@noop {}
  {\bibfield  {journal} {\bibinfo  {journal} {Physical review letters}\
  }\textbf {\bibinfo {volume} {100}},\ \bibinfo {pages} {096407} (\bibinfo
  {year} {2008})}\BibitemShut {NoStop}%
\bibitem [{\citenamefont {Sau}\ \emph {et~al.}(2010)\citenamefont {Sau},
  \citenamefont {Lutchyn}, \citenamefont {Tewari},\ and\ \citenamefont
  {Das~Sarma}}]{sau2010generic}%
  \BibitemOpen
  \bibfield  {author} {\bibinfo {author} {\bibfnamefont {J.~D.}\ \bibnamefont
  {Sau}}, \bibinfo {author} {\bibfnamefont {R.~M.}\ \bibnamefont {Lutchyn}},
  \bibinfo {author} {\bibfnamefont {S.}~\bibnamefont {Tewari}},\ and\ \bibinfo
  {author} {\bibfnamefont {S.}~\bibnamefont {Das~Sarma}},\ }\bibfield  {title}
  {\bibinfo {title} {Generic new platform for topological quantum computation
  using semiconductor heterostructures},\ }\href@noop {} {\bibfield  {journal}
  {\bibinfo  {journal} {Physical review letters}\ }\textbf {\bibinfo {volume}
  {104}},\ \bibinfo {pages} {040502} (\bibinfo {year} {2010})}\BibitemShut
  {NoStop}%
\bibitem [{\citenamefont {Nadj-Perge}\ \emph {et~al.}(2014)\citenamefont
  {Nadj-Perge}, \citenamefont {Drozdov}, \citenamefont {Li}, \citenamefont
  {Chen}, \citenamefont {Jeon}, \citenamefont {Seo}, \citenamefont {MacDonald},
  \citenamefont {Bernevig},\ and\ \citenamefont
  {Yazdani}}]{nadj2014observation}%
  \BibitemOpen
  \bibfield  {author} {\bibinfo {author} {\bibfnamefont {S.}~\bibnamefont
  {Nadj-Perge}}, \bibinfo {author} {\bibfnamefont {I.~K.}\ \bibnamefont
  {Drozdov}}, \bibinfo {author} {\bibfnamefont {J.}~\bibnamefont {Li}},
  \bibinfo {author} {\bibfnamefont {H.}~\bibnamefont {Chen}}, \bibinfo {author}
  {\bibfnamefont {S.}~\bibnamefont {Jeon}}, \bibinfo {author} {\bibfnamefont
  {J.}~\bibnamefont {Seo}}, \bibinfo {author} {\bibfnamefont {A.~H.}\
  \bibnamefont {MacDonald}}, \bibinfo {author} {\bibfnamefont {B.~A.}\
  \bibnamefont {Bernevig}},\ and\ \bibinfo {author} {\bibfnamefont
  {A.}~\bibnamefont {Yazdani}},\ }\bibfield  {title} {\bibinfo {title}
  {Observation of majorana fermions in ferromagnetic atomic chains on a
  superconductor},\ }\href@noop {} {\bibfield  {journal} {\bibinfo  {journal}
  {Science}\ }\textbf {\bibinfo {volume} {346}},\ \bibinfo {pages} {602}
  (\bibinfo {year} {2014})}\BibitemShut {NoStop}%
\bibitem [{\citenamefont {Hao}\ and\ \citenamefont
  {Hu}(2014)}]{hao2014topological}%
  \BibitemOpen
  \bibfield  {author} {\bibinfo {author} {\bibfnamefont {N.}~\bibnamefont
  {Hao}}\ and\ \bibinfo {author} {\bibfnamefont {J.}~\bibnamefont {Hu}},\
  }\bibfield  {title} {\bibinfo {title} {Topological phases in the single-layer
  fese},\ }\href@noop {} {\bibfield  {journal} {\bibinfo  {journal} {Physical
  Review X}\ }\textbf {\bibinfo {volume} {4}},\ \bibinfo {pages} {031053}
  (\bibinfo {year} {2014})}\BibitemShut {NoStop}%
\bibitem [{\citenamefont {Wang}\ \emph {et~al.}(2015)\citenamefont {Wang},
  \citenamefont {Zhang}, \citenamefont {Xu}, \citenamefont {Zeng},
  \citenamefont {Miao}, \citenamefont {Xu}, \citenamefont {Qian}, \citenamefont
  {Weng}, \citenamefont {Richard}, \citenamefont {Fedorov} \emph
  {et~al.}}]{wang2015topological}%
  \BibitemOpen
  \bibfield  {author} {\bibinfo {author} {\bibfnamefont {Z.}~\bibnamefont
  {Wang}}, \bibinfo {author} {\bibfnamefont {P.}~\bibnamefont {Zhang}},
  \bibinfo {author} {\bibfnamefont {G.}~\bibnamefont {Xu}}, \bibinfo {author}
  {\bibfnamefont {L.}~\bibnamefont {Zeng}}, \bibinfo {author} {\bibfnamefont
  {H.}~\bibnamefont {Miao}}, \bibinfo {author} {\bibfnamefont {X.}~\bibnamefont
  {Xu}}, \bibinfo {author} {\bibfnamefont {T.}~\bibnamefont {Qian}}, \bibinfo
  {author} {\bibfnamefont {H.}~\bibnamefont {Weng}}, \bibinfo {author}
  {\bibfnamefont {P.}~\bibnamefont {Richard}}, \bibinfo {author} {\bibfnamefont
  {A.~V.}\ \bibnamefont {Fedorov}}, \emph {et~al.},\ }\bibfield  {title}
  {\bibinfo {title} {Topological nature of the fese 0.5 te 0.5
  superconductor},\ }\href@noop {} {\bibfield  {journal} {\bibinfo  {journal}
  {Physical Review B}\ }\textbf {\bibinfo {volume} {92}},\ \bibinfo {pages}
  {115119} (\bibinfo {year} {2015})}\BibitemShut {NoStop}%
\bibitem [{\citenamefont {Wu}\ \emph {et~al.}(2016)\citenamefont {Wu},
  \citenamefont {Qin}, \citenamefont {Liang}, \citenamefont {Fan},\ and\
  \citenamefont {Hu}}]{wu2016topological}%
  \BibitemOpen
  \bibfield  {author} {\bibinfo {author} {\bibfnamefont {X.}~\bibnamefont
  {Wu}}, \bibinfo {author} {\bibfnamefont {S.}~\bibnamefont {Qin}}, \bibinfo
  {author} {\bibfnamefont {Y.}~\bibnamefont {Liang}}, \bibinfo {author}
  {\bibfnamefont {H.}~\bibnamefont {Fan}},\ and\ \bibinfo {author}
  {\bibfnamefont {J.}~\bibnamefont {Hu}},\ }\bibfield  {title} {\bibinfo
  {title} {Topological characters in fe (te 1- x se x) thin films},\
  }\href@noop {} {\bibfield  {journal} {\bibinfo  {journal} {Physical Review
  B}\ }\textbf {\bibinfo {volume} {93}},\ \bibinfo {pages} {115129} (\bibinfo
  {year} {2016})}\BibitemShut {NoStop}%
\bibitem [{\citenamefont {Hao}\ and\ \citenamefont
  {Hu}(2019)}]{hao2019topological}%
  \BibitemOpen
  \bibfield  {author} {\bibinfo {author} {\bibfnamefont {N.}~\bibnamefont
  {Hao}}\ and\ \bibinfo {author} {\bibfnamefont {J.}~\bibnamefont {Hu}},\
  }\bibfield  {title} {\bibinfo {title} {Topological quantum states of matter
  in iron-based superconductors: from concept to material realization},\
  }\href@noop {} {\bibfield  {journal} {\bibinfo  {journal} {National Science
  Review}\ }\textbf {\bibinfo {volume} {6}},\ \bibinfo {pages} {213} (\bibinfo
  {year} {2019})}\BibitemShut {NoStop}%
\bibitem [{\citenamefont {Wu}\ \emph {et~al.}(2015)\citenamefont {Wu},
  \citenamefont {Qin}, \citenamefont {Liang}, \citenamefont {Le}, \citenamefont
  {Fan},\ and\ \citenamefont {Hu}}]{wu2015cafeas}%
  \BibitemOpen
  \bibfield  {author} {\bibinfo {author} {\bibfnamefont {X.}~\bibnamefont
  {Wu}}, \bibinfo {author} {\bibfnamefont {S.}~\bibnamefont {Qin}}, \bibinfo
  {author} {\bibfnamefont {Y.}~\bibnamefont {Liang}}, \bibinfo {author}
  {\bibfnamefont {C.}~\bibnamefont {Le}}, \bibinfo {author} {\bibfnamefont
  {H.}~\bibnamefont {Fan}},\ and\ \bibinfo {author} {\bibfnamefont
  {J.}~\bibnamefont {Hu}},\ }\bibfield  {title} {\bibinfo {title} {Cafeas 2: A
  staggered intercalation of quantum spin hall and high-temperature
  superconductivity},\ }\href@noop {} {\bibfield  {journal} {\bibinfo
  {journal} {Physical Review B}\ }\textbf {\bibinfo {volume} {91}},\ \bibinfo
  {pages} {081111} (\bibinfo {year} {2015})}\BibitemShut {NoStop}%
\bibitem [{\citenamefont {Xu}\ \emph {et~al.}(2016)\citenamefont {Xu},
  \citenamefont {Lian}, \citenamefont {Tang}, \citenamefont {Qi},\ and\
  \citenamefont {Zhang}}]{xu2016topological}%
  \BibitemOpen
  \bibfield  {author} {\bibinfo {author} {\bibfnamefont {G.}~\bibnamefont
  {Xu}}, \bibinfo {author} {\bibfnamefont {B.}~\bibnamefont {Lian}}, \bibinfo
  {author} {\bibfnamefont {P.}~\bibnamefont {Tang}}, \bibinfo {author}
  {\bibfnamefont {X.-L.}\ \bibnamefont {Qi}},\ and\ \bibinfo {author}
  {\bibfnamefont {S.-C.}\ \bibnamefont {Zhang}},\ }\bibfield  {title} {\bibinfo
  {title} {Topological superconductivity on the surface of fe-based
  superconductors},\ }\href@noop {} {\bibfield  {journal} {\bibinfo  {journal}
  {Physical review letters}\ }\textbf {\bibinfo {volume} {117}},\ \bibinfo
  {pages} {047001} (\bibinfo {year} {2016})}\BibitemShut {NoStop}%
\bibitem [{\citenamefont {Chang}\ \emph {et~al.}(2016)\citenamefont {Chang},
  \citenamefont {Chen}, \citenamefont {Bian}, \citenamefont {Huang},
  \citenamefont {Zheng}, \citenamefont {Neupert}, \citenamefont {Sankar},
  \citenamefont {Xu}, \citenamefont {Belopolski}, \citenamefont {Chang} \emph
  {et~al.}}]{chang2016topological}%
  \BibitemOpen
  \bibfield  {author} {\bibinfo {author} {\bibfnamefont {T.-R.}\ \bibnamefont
  {Chang}}, \bibinfo {author} {\bibfnamefont {P.-J.}\ \bibnamefont {Chen}},
  \bibinfo {author} {\bibfnamefont {G.}~\bibnamefont {Bian}}, \bibinfo {author}
  {\bibfnamefont {S.-M.}\ \bibnamefont {Huang}}, \bibinfo {author}
  {\bibfnamefont {H.}~\bibnamefont {Zheng}}, \bibinfo {author} {\bibfnamefont
  {T.}~\bibnamefont {Neupert}}, \bibinfo {author} {\bibfnamefont
  {R.}~\bibnamefont {Sankar}}, \bibinfo {author} {\bibfnamefont {S.-Y.}\
  \bibnamefont {Xu}}, \bibinfo {author} {\bibfnamefont {I.}~\bibnamefont
  {Belopolski}}, \bibinfo {author} {\bibfnamefont {G.}~\bibnamefont {Chang}},
  \emph {et~al.},\ }\bibfield  {title} {\bibinfo {title} {Topological dirac
  surface states and superconducting pairing correlations in pbtase 2},\
  }\href@noop {} {\bibfield  {journal} {\bibinfo  {journal} {Physical Review
  B}\ }\textbf {\bibinfo {volume} {93}},\ \bibinfo {pages} {245130} (\bibinfo
  {year} {2016})}\BibitemShut {NoStop}%
\bibitem [{\citenamefont {Guan}\ \emph {et~al.}(2016)\citenamefont {Guan},
  \citenamefont {Chen}, \citenamefont {Chu}, \citenamefont {Sankar},
  \citenamefont {Chou}, \citenamefont {Jeng}, \citenamefont {Chang},\ and\
  \citenamefont {Chuang}}]{guan2016superconducting}%
  \BibitemOpen
  \bibfield  {author} {\bibinfo {author} {\bibfnamefont {S.-Y.}\ \bibnamefont
  {Guan}}, \bibinfo {author} {\bibfnamefont {P.-J.}\ \bibnamefont {Chen}},
  \bibinfo {author} {\bibfnamefont {M.-W.}\ \bibnamefont {Chu}}, \bibinfo
  {author} {\bibfnamefont {R.}~\bibnamefont {Sankar}}, \bibinfo {author}
  {\bibfnamefont {F.}~\bibnamefont {Chou}}, \bibinfo {author} {\bibfnamefont
  {H.-T.}\ \bibnamefont {Jeng}}, \bibinfo {author} {\bibfnamefont {C.-S.}\
  \bibnamefont {Chang}},\ and\ \bibinfo {author} {\bibfnamefont {T.-M.}\
  \bibnamefont {Chuang}},\ }\bibfield  {title} {\bibinfo {title}
  {Superconducting topological surface states in the noncentrosymmetric bulk
  superconductor pbtase2},\ }\href@noop {} {\bibfield  {journal} {\bibinfo
  {journal} {Science advances}\ }\textbf {\bibinfo {volume} {2}},\ \bibinfo
  {pages} {e1600894} (\bibinfo {year} {2016})}\BibitemShut {NoStop}%
\bibitem [{\citenamefont {Jiao}\ \emph {et~al.}(2020)\citenamefont {Jiao},
  \citenamefont {Howard}, \citenamefont {Ran}, \citenamefont {Wang},
  \citenamefont {Rodriguez}, \citenamefont {Sigrist}, \citenamefont {Wang},
  \citenamefont {Butch},\ and\ \citenamefont {Madhavan}}]{jiao2020chiral}%
  \BibitemOpen
  \bibfield  {author} {\bibinfo {author} {\bibfnamefont {L.}~\bibnamefont
  {Jiao}}, \bibinfo {author} {\bibfnamefont {S.}~\bibnamefont {Howard}},
  \bibinfo {author} {\bibfnamefont {S.}~\bibnamefont {Ran}}, \bibinfo {author}
  {\bibfnamefont {Z.}~\bibnamefont {Wang}}, \bibinfo {author} {\bibfnamefont
  {J.~O.}\ \bibnamefont {Rodriguez}}, \bibinfo {author} {\bibfnamefont
  {M.}~\bibnamefont {Sigrist}}, \bibinfo {author} {\bibfnamefont
  {Z.}~\bibnamefont {Wang}}, \bibinfo {author} {\bibfnamefont {N.~P.}\
  \bibnamefont {Butch}},\ and\ \bibinfo {author} {\bibfnamefont
  {V.}~\bibnamefont {Madhavan}},\ }\bibfield  {title} {\bibinfo {title} {Chiral
  superconductivity in heavy-fermion metal ute2},\ }\href@noop {} {\bibfield
  {journal} {\bibinfo  {journal} {Nature}\ }\textbf {\bibinfo {volume} {579}},\
  \bibinfo {pages} {523} (\bibinfo {year} {2020})}\BibitemShut {NoStop}%
\bibitem [{\citenamefont {Aoki}\ \emph {et~al.}(2019)\citenamefont {Aoki},
  \citenamefont {Nakamura}, \citenamefont {Honda}, \citenamefont {Li},
  \citenamefont {Homma}, \citenamefont {Shimizu}, \citenamefont {Sato},
  \citenamefont {Knebel}, \citenamefont {Brison}, \citenamefont {Pourret} \emph
  {et~al.}}]{aoki2019unconventional}%
  \BibitemOpen
  \bibfield  {author} {\bibinfo {author} {\bibfnamefont {D.}~\bibnamefont
  {Aoki}}, \bibinfo {author} {\bibfnamefont {A.}~\bibnamefont {Nakamura}},
  \bibinfo {author} {\bibfnamefont {F.}~\bibnamefont {Honda}}, \bibinfo
  {author} {\bibfnamefont {D.}~\bibnamefont {Li}}, \bibinfo {author}
  {\bibfnamefont {Y.}~\bibnamefont {Homma}}, \bibinfo {author} {\bibfnamefont
  {Y.}~\bibnamefont {Shimizu}}, \bibinfo {author} {\bibfnamefont {Y.~J.}\
  \bibnamefont {Sato}}, \bibinfo {author} {\bibfnamefont {G.}~\bibnamefont
  {Knebel}}, \bibinfo {author} {\bibfnamefont {J.-P.}\ \bibnamefont {Brison}},
  \bibinfo {author} {\bibfnamefont {A.}~\bibnamefont {Pourret}}, \emph
  {et~al.},\ }\bibfield  {title} {\bibinfo {title} {Unconventional
  superconductivity in heavy fermion ute2},\ }\href@noop {} {\bibfield
  {journal} {\bibinfo  {journal} {journal of the physical society of japan}\
  }\textbf {\bibinfo {volume} {88}},\ \bibinfo {pages} {043702} (\bibinfo
  {year} {2019})}\BibitemShut {NoStop}%
\bibitem [{\citenamefont {Aoki}\ \emph {et~al.}(2022)\citenamefont {Aoki},
  \citenamefont {Brison}, \citenamefont {Flouquet}, \citenamefont {Ishida},
  \citenamefont {Knebel}, \citenamefont {Tokunaga},\ and\ \citenamefont
  {Yanase}}]{aoki2022unconventional}%
  \BibitemOpen
  \bibfield  {author} {\bibinfo {author} {\bibfnamefont {D.}~\bibnamefont
  {Aoki}}, \bibinfo {author} {\bibfnamefont {J.-P.}\ \bibnamefont {Brison}},
  \bibinfo {author} {\bibfnamefont {J.}~\bibnamefont {Flouquet}}, \bibinfo
  {author} {\bibfnamefont {K.}~\bibnamefont {Ishida}}, \bibinfo {author}
  {\bibfnamefont {G.}~\bibnamefont {Knebel}}, \bibinfo {author} {\bibfnamefont
  {Y.}~\bibnamefont {Tokunaga}},\ and\ \bibinfo {author} {\bibfnamefont
  {Y.}~\bibnamefont {Yanase}},\ }\bibfield  {title} {\bibinfo {title}
  {Unconventional superconductivity in ute2},\ }\href@noop {} {\bibfield
  {journal} {\bibinfo  {journal} {Journal of Physics: Condensed Matter}\
  }\textbf {\bibinfo {volume} {34}},\ \bibinfo {pages} {243002} (\bibinfo
  {year} {2022})}\BibitemShut {NoStop}%
\bibitem [{\citenamefont {Bauer}\ \emph {et~al.}(2004)\citenamefont {Bauer},
  \citenamefont {Hilscher}, \citenamefont {Michor}, \citenamefont {Paul},
  \citenamefont {Scheidt}, \citenamefont {Gribanov}, \citenamefont {Seropegin},
  \citenamefont {No{\"e}l}, \citenamefont {Sigrist},\ and\ \citenamefont
  {Rogl}}]{bauer2004heavy}%
  \BibitemOpen
  \bibfield  {author} {\bibinfo {author} {\bibfnamefont {E.}~\bibnamefont
  {Bauer}}, \bibinfo {author} {\bibfnamefont {G.}~\bibnamefont {Hilscher}},
  \bibinfo {author} {\bibfnamefont {H.}~\bibnamefont {Michor}}, \bibinfo
  {author} {\bibfnamefont {C.}~\bibnamefont {Paul}}, \bibinfo {author}
  {\bibfnamefont {E.-W.}\ \bibnamefont {Scheidt}}, \bibinfo {author}
  {\bibfnamefont {A.}~\bibnamefont {Gribanov}}, \bibinfo {author}
  {\bibfnamefont {Y.}~\bibnamefont {Seropegin}}, \bibinfo {author}
  {\bibfnamefont {H.}~\bibnamefont {No{\"e}l}}, \bibinfo {author}
  {\bibfnamefont {M.}~\bibnamefont {Sigrist}},\ and\ \bibinfo {author}
  {\bibfnamefont {P.}~\bibnamefont {Rogl}},\ }\bibfield  {title} {\bibinfo
  {title} {Heavy fermion superconductivity and magnetic order in
  noncentrosymmetric c e p t 3 s i},\ }\href@noop {} {\bibfield  {journal}
  {\bibinfo  {journal} {Physical review letters}\ }\textbf {\bibinfo {volume}
  {92}},\ \bibinfo {pages} {027003} (\bibinfo {year} {2004})}\BibitemShut
  {NoStop}%
\bibitem [{\citenamefont {Smidman}\ \emph {et~al.}(2017)\citenamefont
  {Smidman}, \citenamefont {Salamon}, \citenamefont {Yuan},\ and\ \citenamefont
  {Agterberg}}]{smidman2017}%
  \BibitemOpen
  \bibfield  {author} {\bibinfo {author} {\bibfnamefont {M.}~\bibnamefont
  {Smidman}}, \bibinfo {author} {\bibfnamefont {M.}~\bibnamefont {Salamon}},
  \bibinfo {author} {\bibfnamefont {H.}~\bibnamefont {Yuan}},\ and\ \bibinfo
  {author} {\bibfnamefont {D.}~\bibnamefont {Agterberg}},\ }\bibfield  {title}
  {\bibinfo {title} {Superconductivity and spin--orbit coupling in
  non-centrosymmetric materials: a review},\ }\href@noop {} {\bibfield
  {journal} {\bibinfo  {journal} {Reports on Progress in Physics}\ }\textbf
  {\bibinfo {volume} {80}},\ \bibinfo {pages} {036501} (\bibinfo {year}
  {2017})}\BibitemShut {NoStop}%
\bibitem [{\citenamefont {Krupka}\ \emph {et~al.}(1969)\citenamefont {Krupka},
  \citenamefont {Giorgi}, \citenamefont {Krikorian},\ and\ \citenamefont
  {Szklarz}}]{krupka1969high}%
  \BibitemOpen
  \bibfield  {author} {\bibinfo {author} {\bibfnamefont {M.}~\bibnamefont
  {Krupka}}, \bibinfo {author} {\bibfnamefont {A.}~\bibnamefont {Giorgi}},
  \bibinfo {author} {\bibfnamefont {N.}~\bibnamefont {Krikorian}},\ and\
  \bibinfo {author} {\bibfnamefont {E.}~\bibnamefont {Szklarz}},\ }\bibfield
  {title} {\bibinfo {title} {High pressure synthesis and superconducting
  properties of yttrium sesquicarbide},\ }\href@noop {} {\bibfield  {journal}
  {\bibinfo  {journal} {Journal of the Less Common Metals}\ }\textbf {\bibinfo
  {volume} {17}},\ \bibinfo {pages} {91} (\bibinfo {year} {1969})}\BibitemShut
  {NoStop}%
\bibitem [{\citenamefont {Bao}\ \emph {et~al.}(2015)\citenamefont {Bao},
  \citenamefont {Liu}, \citenamefont {Ma}, \citenamefont {Meng}, \citenamefont
  {Tang}, \citenamefont {Sun}, \citenamefont {Zhai}, \citenamefont {Jiang},
  \citenamefont {Bai}, \citenamefont {Feng} \emph
  {et~al.}}]{bao2015superconductivity}%
  \BibitemOpen
  \bibfield  {author} {\bibinfo {author} {\bibfnamefont {J.-K.}\ \bibnamefont
  {Bao}}, \bibinfo {author} {\bibfnamefont {J.-Y.}\ \bibnamefont {Liu}},
  \bibinfo {author} {\bibfnamefont {C.-W.}\ \bibnamefont {Ma}}, \bibinfo
  {author} {\bibfnamefont {Z.-H.}\ \bibnamefont {Meng}}, \bibinfo {author}
  {\bibfnamefont {Z.-T.}\ \bibnamefont {Tang}}, \bibinfo {author}
  {\bibfnamefont {Y.-L.}\ \bibnamefont {Sun}}, \bibinfo {author} {\bibfnamefont
  {H.-F.}\ \bibnamefont {Zhai}}, \bibinfo {author} {\bibfnamefont
  {H.}~\bibnamefont {Jiang}}, \bibinfo {author} {\bibfnamefont
  {H.}~\bibnamefont {Bai}}, \bibinfo {author} {\bibfnamefont {C.-M.}\
  \bibnamefont {Feng}}, \emph {et~al.},\ }\bibfield  {title} {\bibinfo {title}
  {Superconductivity in quasi-one-dimensional k 2 cr 3 as 3 with significant
  electron correlations},\ }\href@noop {} {\bibfield  {journal} {\bibinfo
  {journal} {Physical Review X}\ }\textbf {\bibinfo {volume} {5}},\ \bibinfo
  {pages} {011013} (\bibinfo {year} {2015})}\BibitemShut {NoStop}%
\bibitem [{\citenamefont {Tang}\ \emph
  {et~al.}(2015{\natexlab{a}})\citenamefont {Tang}, \citenamefont {Bao},
  \citenamefont {Liu}, \citenamefont {Sun}, \citenamefont {Ablimit},
  \citenamefont {Zhai}, \citenamefont {Jiang}, \citenamefont {Feng},
  \citenamefont {Xu},\ and\ \citenamefont {Cao}}]{tang2015unconventional}%
  \BibitemOpen
  \bibfield  {author} {\bibinfo {author} {\bibfnamefont {Z.-T.}\ \bibnamefont
  {Tang}}, \bibinfo {author} {\bibfnamefont {J.-K.}\ \bibnamefont {Bao}},
  \bibinfo {author} {\bibfnamefont {Y.}~\bibnamefont {Liu}}, \bibinfo {author}
  {\bibfnamefont {Y.-L.}\ \bibnamefont {Sun}}, \bibinfo {author} {\bibfnamefont
  {A.}~\bibnamefont {Ablimit}}, \bibinfo {author} {\bibfnamefont {H.-F.}\
  \bibnamefont {Zhai}}, \bibinfo {author} {\bibfnamefont {H.}~\bibnamefont
  {Jiang}}, \bibinfo {author} {\bibfnamefont {C.-M.}\ \bibnamefont {Feng}},
  \bibinfo {author} {\bibfnamefont {Z.-A.}\ \bibnamefont {Xu}},\ and\ \bibinfo
  {author} {\bibfnamefont {G.-H.}\ \bibnamefont {Cao}},\ }\bibfield  {title}
  {\bibinfo {title} {Unconventional superconductivity in quasi-one-dimensional
  rb 2 cr 3 as 3},\ }\href@noop {} {\bibfield  {journal} {\bibinfo  {journal}
  {Physical Review B}\ }\textbf {\bibinfo {volume} {91}},\ \bibinfo {pages}
  {020506} (\bibinfo {year} {2015}{\natexlab{a}})}\BibitemShut {NoStop}%
\bibitem [{\citenamefont {Tang}\ \emph
  {et~al.}(2015{\natexlab{b}})\citenamefont {Tang}, \citenamefont {Bao},
  \citenamefont {Wang}, \citenamefont {Bai}, \citenamefont {Jiang},
  \citenamefont {Liu}, \citenamefont {Zhai}, \citenamefont {Feng},
  \citenamefont {Xu},\ and\ \citenamefont {Cao}}]{tang2015superconductivity}%
  \BibitemOpen
  \bibfield  {author} {\bibinfo {author} {\bibfnamefont {Z.-T.}\ \bibnamefont
  {Tang}}, \bibinfo {author} {\bibfnamefont {J.-K.}\ \bibnamefont {Bao}},
  \bibinfo {author} {\bibfnamefont {Z.}~\bibnamefont {Wang}}, \bibinfo {author}
  {\bibfnamefont {H.}~\bibnamefont {Bai}}, \bibinfo {author} {\bibfnamefont
  {H.}~\bibnamefont {Jiang}}, \bibinfo {author} {\bibfnamefont
  {Y.}~\bibnamefont {Liu}}, \bibinfo {author} {\bibfnamefont {H.-F.}\
  \bibnamefont {Zhai}}, \bibinfo {author} {\bibfnamefont {C.-M.}\ \bibnamefont
  {Feng}}, \bibinfo {author} {\bibfnamefont {Z.-A.}\ \bibnamefont {Xu}},\ and\
  \bibinfo {author} {\bibfnamefont {G.-H.}\ \bibnamefont {Cao}},\ }\bibfield
  {title} {\bibinfo {title} {Superconductivity in quasi-one-dimensional cs 2 cr
  3 as 3 with large interchain distance},\ }\href@noop {} {\bibfield  {journal}
  {\bibinfo  {journal} {Science China Materials}\ }\textbf {\bibinfo {volume}
  {58}},\ \bibinfo {pages} {16} (\bibinfo {year}
  {2015}{\natexlab{b}})}\BibitemShut {NoStop}%
\bibitem [{\citenamefont {Gotlieb}\ \emph {et~al.}(2018)\citenamefont
  {Gotlieb}, \citenamefont {Lin}, \citenamefont {Serbyn}, \citenamefont
  {Zhang}, \citenamefont {Smallwood}, \citenamefont {Jozwiak}, \citenamefont
  {Eisaki}, \citenamefont {Hussain}, \citenamefont {Vishwanath},\ and\
  \citenamefont {Lanzara}}]{gotlieb2018revealing}%
  \BibitemOpen
  \bibfield  {author} {\bibinfo {author} {\bibfnamefont {K.}~\bibnamefont
  {Gotlieb}}, \bibinfo {author} {\bibfnamefont {C.-Y.}\ \bibnamefont {Lin}},
  \bibinfo {author} {\bibfnamefont {M.}~\bibnamefont {Serbyn}}, \bibinfo
  {author} {\bibfnamefont {W.}~\bibnamefont {Zhang}}, \bibinfo {author}
  {\bibfnamefont {C.~L.}\ \bibnamefont {Smallwood}}, \bibinfo {author}
  {\bibfnamefont {C.}~\bibnamefont {Jozwiak}}, \bibinfo {author} {\bibfnamefont
  {H.}~\bibnamefont {Eisaki}}, \bibinfo {author} {\bibfnamefont
  {Z.}~\bibnamefont {Hussain}}, \bibinfo {author} {\bibfnamefont
  {A.}~\bibnamefont {Vishwanath}},\ and\ \bibinfo {author} {\bibfnamefont
  {A.}~\bibnamefont {Lanzara}},\ }\bibfield  {title} {\bibinfo {title}
  {Revealing hidden spin-momentum locking in a high-temperature cuprate
  superconductor},\ }\href@noop {} {\bibfield  {journal} {\bibinfo  {journal}
  {Science}\ }\textbf {\bibinfo {volume} {362}},\ \bibinfo {pages} {1271}
  (\bibinfo {year} {2018})}\BibitemShut {NoStop}%
\bibitem [{\citenamefont {Borisenko}\ \emph {et~al.}(2016)\citenamefont
  {Borisenko}, \citenamefont {Evtushinsky}, \citenamefont {Liu}, \citenamefont
  {Morozov}, \citenamefont {Kappenberger}, \citenamefont {Wurmehl},
  \citenamefont {B{\"u}chner}, \citenamefont {Yaresko}, \citenamefont {Kim},
  \citenamefont {Hoesch} \emph {et~al.}}]{borisenko2016}%
  \BibitemOpen
  \bibfield  {author} {\bibinfo {author} {\bibfnamefont {S.}~\bibnamefont
  {Borisenko}}, \bibinfo {author} {\bibfnamefont {D.}~\bibnamefont
  {Evtushinsky}}, \bibinfo {author} {\bibfnamefont {Z.-H.}\ \bibnamefont
  {Liu}}, \bibinfo {author} {\bibfnamefont {I.}~\bibnamefont {Morozov}},
  \bibinfo {author} {\bibfnamefont {R.}~\bibnamefont {Kappenberger}}, \bibinfo
  {author} {\bibfnamefont {S.}~\bibnamefont {Wurmehl}}, \bibinfo {author}
  {\bibfnamefont {B.}~\bibnamefont {B{\"u}chner}}, \bibinfo {author}
  {\bibfnamefont {A.}~\bibnamefont {Yaresko}}, \bibinfo {author} {\bibfnamefont
  {T.}~\bibnamefont {Kim}}, \bibinfo {author} {\bibfnamefont {M.}~\bibnamefont
  {Hoesch}}, \emph {et~al.},\ }\bibfield  {title} {\bibinfo {title} {Direct
  observation of spin--orbit coupling in iron-based superconductors},\
  }\href@noop {} {\bibfield  {journal} {\bibinfo  {journal} {Nature Physics}\
  }\textbf {\bibinfo {volume} {12}},\ \bibinfo {pages} {311} (\bibinfo {year}
  {2016})}\BibitemShut {NoStop}%
\bibitem [{\citenamefont {Greco}\ and\ \citenamefont
  {Schnyder}(2018)}]{greco2018mechanism}%
  \BibitemOpen
  \bibfield  {author} {\bibinfo {author} {\bibfnamefont {A.}~\bibnamefont
  {Greco}}\ and\ \bibinfo {author} {\bibfnamefont {A.~P.}\ \bibnamefont
  {Schnyder}},\ }\bibfield  {title} {\bibinfo {title} {Mechanism for
  unconventional superconductivity in the hole-doped rashba-hubbard model},\
  }\href@noop {} {\bibfield  {journal} {\bibinfo  {journal} {Physical Review
  Letters}\ }\textbf {\bibinfo {volume} {120}},\ \bibinfo {pages} {177002}
  (\bibinfo {year} {2018})}\BibitemShut {NoStop}%
\bibitem [{\citenamefont {Nogaki}\ and\ \citenamefont
  {Yanase}(2020)}]{nogaki2020}%
  \BibitemOpen
  \bibfield  {author} {\bibinfo {author} {\bibfnamefont {K.}~\bibnamefont
  {Nogaki}}\ and\ \bibinfo {author} {\bibfnamefont {Y.}~\bibnamefont
  {Yanase}},\ }\bibfield  {title} {\bibinfo {title} {Strongly parity-mixed
  superconductivity in the rashba-hubbard model},\ }\href@noop {} {\bibfield
  {journal} {\bibinfo  {journal} {Physical Review B}\ }\textbf {\bibinfo
  {volume} {102}},\ \bibinfo {pages} {165114} (\bibinfo {year}
  {2020})}\BibitemShut {NoStop}%
\bibitem [{\citenamefont {Greco}\ \emph {et~al.}(2020)\citenamefont {Greco},
  \citenamefont {Bejas},\ and\ \citenamefont {Schnyder}}]{greco2020}%
  \BibitemOpen
  \bibfield  {author} {\bibinfo {author} {\bibfnamefont {A.}~\bibnamefont
  {Greco}}, \bibinfo {author} {\bibfnamefont {M.}~\bibnamefont {Bejas}},\ and\
  \bibinfo {author} {\bibfnamefont {A.~P.}\ \bibnamefont {Schnyder}},\
  }\bibfield  {title} {\bibinfo {title} {Ferromagnetic fluctuations in the
  rashba-hubbard model},\ }\href@noop {} {\bibfield  {journal} {\bibinfo
  {journal} {Physical Review B}\ }\textbf {\bibinfo {volume} {101}},\ \bibinfo
  {pages} {174420} (\bibinfo {year} {2020})}\BibitemShut {NoStop}%
\bibitem [{\citenamefont {Bonetti}\ \emph {et~al.}(2024)\citenamefont
  {Bonetti}, \citenamefont {Chakraborty}, \citenamefont {Wu},\ and\
  \citenamefont {Schnyder}}]{bonetti2024}%
  \BibitemOpen
  \bibfield  {author} {\bibinfo {author} {\bibfnamefont {P.~M.}\ \bibnamefont
  {Bonetti}}, \bibinfo {author} {\bibfnamefont {D.}~\bibnamefont
  {Chakraborty}}, \bibinfo {author} {\bibfnamefont {X.}~\bibnamefont {Wu}},\
  and\ \bibinfo {author} {\bibfnamefont {A.~P.}\ \bibnamefont {Schnyder}},\
  }\bibfield  {title} {\bibinfo {title} {Interaction-driven first-order and
  higher-order topological superconductivity},\ }\href@noop {} {\bibfield
  {journal} {\bibinfo  {journal} {Physical Review B}\ }\textbf {\bibinfo
  {volume} {109}},\ \bibinfo {pages} {L180509} (\bibinfo {year}
  {2024})}\BibitemShut {NoStop}%
\bibitem [{\citenamefont {Huang}\ \emph {et~al.}(2024)\citenamefont {Huang},
  \citenamefont {Xiao}, \citenamefont {Song},\ and\ \citenamefont
  {Hao}}]{huang2024}%
  \BibitemOpen
  \bibfield  {author} {\bibinfo {author} {\bibfnamefont {X.}~\bibnamefont
  {Huang}}, \bibinfo {author} {\bibfnamefont {Y.}~\bibnamefont {Xiao}},
  \bibinfo {author} {\bibfnamefont {R.}~\bibnamefont {Song}},\ and\ \bibinfo
  {author} {\bibfnamefont {N.}~\bibnamefont {Hao}},\ }\bibfield  {title}
  {\bibinfo {title} {Generic model with unconventional rashba bands and giant
  spin galvanic effect},\ }\href@noop {} {\bibfield  {journal} {\bibinfo
  {journal} {Physical Review B}\ }\textbf {\bibinfo {volume} {109}},\ \bibinfo
  {pages} {195419} (\bibinfo {year} {2024})}\BibitemShut {NoStop}%
\bibitem [{\citenamefont {Wang}\ \emph {et~al.}(2024)\citenamefont {Wang},
  \citenamefont {Li}, \citenamefont {Huang}, \citenamefont {Wang},
  \citenamefont {Song},\ and\ \citenamefont {Hao}}]{wang2024}%
  \BibitemOpen
  \bibfield  {author} {\bibinfo {author} {\bibfnamefont {R.}~\bibnamefont
  {Wang}}, \bibinfo {author} {\bibfnamefont {J.}~\bibnamefont {Li}}, \bibinfo
  {author} {\bibfnamefont {X.}~\bibnamefont {Huang}}, \bibinfo {author}
  {\bibfnamefont {L.}~\bibnamefont {Wang}}, \bibinfo {author} {\bibfnamefont
  {R.}~\bibnamefont {Song}},\ and\ \bibinfo {author} {\bibfnamefont
  {N.}~\bibnamefont {Hao}},\ }\bibfield  {title} {\bibinfo {title}
  {Superconductivity in two-dimensional systems with unconventional rashba
  bands},\ }\href@noop {} {\bibfield  {journal} {\bibinfo  {journal} {Physical
  Review B}\ }\textbf {\bibinfo {volume} {110}},\ \bibinfo {pages} {134517}
  (\bibinfo {year} {2024})}\BibitemShut {NoStop}%
\bibitem [{\citenamefont {Wang}\ \emph {et~al.}(2025)\citenamefont {Wang},
  \citenamefont {Zhang},\ and\ \citenamefont {Hao}}]{wang2025finite}%
  \BibitemOpen
  \bibfield  {author} {\bibinfo {author} {\bibfnamefont {R.}~\bibnamefont
  {Wang}}, \bibinfo {author} {\bibfnamefont {S.-B.}\ \bibnamefont {Zhang}},\
  and\ \bibinfo {author} {\bibfnamefont {N.}~\bibnamefont {Hao}},\ }\bibfield
  {title} {\bibinfo {title} {Finite-momentum pairing state in unconventional
  rashba systems},\ }\href@noop {} {\bibfield  {journal} {\bibinfo  {journal}
  {Physical Review B}\ }\textbf {\bibinfo {volume} {111}},\ \bibinfo {pages}
  {L100506} (\bibinfo {year} {2025})}\BibitemShut {NoStop}%
\bibitem [{\citenamefont {Huang}\ \emph {et~al.}(2025)\citenamefont {Huang},
  \citenamefont {Xiao}, \citenamefont {Song},\ and\ \citenamefont
  {Hao}}]{huang2024-1}%
  \BibitemOpen
  \bibfield  {author} {\bibinfo {author} {\bibfnamefont {X.}~\bibnamefont
  {Huang}}, \bibinfo {author} {\bibfnamefont {Y.}~\bibnamefont {Xiao}},
  \bibinfo {author} {\bibfnamefont {R.}~\bibnamefont {Song}},\ and\ \bibinfo
  {author} {\bibfnamefont {N.}~\bibnamefont {Hao}},\ }\bibfield  {title}
  {\bibinfo {title} {Erratum: Generic model with unconventional rashba bands
  and giant spin galvanic effect [phys. rev. b 109, 195419 (2024)]},\ }\href
  {https://doi.org/10.1103/tvq4-119x} {\bibfield  {journal} {\bibinfo
  {journal} {Phys. Rev. B}\ }\textbf {\bibinfo {volume} {112}},\ \bibinfo
  {pages} {119902} (\bibinfo {year} {2025})}\BibitemShut {NoStop}%
\bibitem [{\citenamefont {Kane}\ and\ \citenamefont
  {Mele}(2005)}]{PhysRevLett.95.146802}%
  \BibitemOpen
  \bibfield  {author} {\bibinfo {author} {\bibfnamefont {C.~L.}\ \bibnamefont
  {Kane}}\ and\ \bibinfo {author} {\bibfnamefont {E.~J.}\ \bibnamefont
  {Mele}},\ }\bibfield  {title} {\bibinfo {title} {${Z}_{2}$ topological order
  and the quantum spin hall effect},\ }\href
  {https://doi.org/10.1103/PhysRevLett.95.146802} {\bibfield  {journal}
  {\bibinfo  {journal} {Phys. Rev. Lett.}\ }\textbf {\bibinfo {volume} {95}},\
  \bibinfo {pages} {146802} (\bibinfo {year} {2005})}\BibitemShut {NoStop}%
\bibitem [{not()}]{note1}%
  \BibitemOpen
  \bibfield  {title} {\bibinfo {title} {Here, we use anisotropic intra-orbital
  rsoc $\tau^3$ term to replace the inter-orbital rsoc $\tau^1$ term used in
  refs. \cite{huang2024, wang2024, wang2025finite}. note that these two cases
  are equivalent and can transform in to each other through an unitary operator
  $(\tau^0+i\tau^2)/\sqrt{2}$. here, for convenience to define the model on
  square lattice, we adopt the former case},\ }\href@noop {} {\ }\BibitemShut
  {NoStop}%
\bibitem [{\citenamefont {Berk}\ and\ \citenamefont
  {Schrieffer}(1966)}]{berk1966}%
  \BibitemOpen
  \bibfield  {author} {\bibinfo {author} {\bibfnamefont {N.}~\bibnamefont
  {Berk}}\ and\ \bibinfo {author} {\bibfnamefont {J.}~\bibnamefont
  {Schrieffer}},\ }\bibfield  {title} {\bibinfo {title} {Effect of
  ferromagnetic spin correlations on superconductivity},\ }\href@noop {}
  {\bibfield  {journal} {\bibinfo  {journal} {Physical Review Letters}\
  }\textbf {\bibinfo {volume} {17}},\ \bibinfo {pages} {433} (\bibinfo {year}
  {1966})}\BibitemShut {NoStop}%
\bibitem [{\citenamefont {Emery}(1964)}]{emery1964}%
  \BibitemOpen
  \bibfield  {author} {\bibinfo {author} {\bibfnamefont {V.}~\bibnamefont
  {Emery}},\ }\bibfield  {title} {\bibinfo {title} {Theories of liquid helium
  three},\ }\href@noop {} {\bibfield  {journal} {\bibinfo  {journal} {Annals of
  Physics}\ }\textbf {\bibinfo {volume} {28}},\ \bibinfo {pages} {1} (\bibinfo
  {year} {1964})}\BibitemShut {NoStop}%
\bibitem [{\citenamefont {Scalapino}(2012)}]{scalapino2012}%
  \BibitemOpen
  \bibfield  {author} {\bibinfo {author} {\bibfnamefont {D.~J.}\ \bibnamefont
  {Scalapino}},\ }\bibfield  {title} {\bibinfo {title} {A common thread: The
  pairing interaction for unconventional superconductors},\ }\href@noop {}
  {\bibfield  {journal} {\bibinfo  {journal} {Reviews of Modern Physics}\
  }\textbf {\bibinfo {volume} {84}},\ \bibinfo {pages} {1383} (\bibinfo {year}
  {2012})}\BibitemShut {NoStop}%
\bibitem [{\citenamefont {Markiewicz}\ \emph {et~al.}(2010{\natexlab{b}})\citenamefont
	{Markiewicz}, \citenamefont {Lorenzana},\ and\ \citenamefont {Seibold}, \citenamefont {Bansil}}]
	{PhysRevB.81.014509}%
	\BibitemOpen
	\bibfield  {author} {\bibinfo {author} {\bibfnamefont {R.S.}~\bibnamefont
		{Markiewicz}}, \bibinfo {author} {\bibfnamefont {J.}\ \bibnamefont {Lorenzana}},\ and\ \bibinfo
	{author} {\bibfnamefont {G.}~\bibnamefont {Seibold}}, \bibinfo {author}
	{\bibfnamefont {A.}~\bibnamefont {Bansil}},\ }\bibfield  {title} {\bibinfo
	{title} {Gutzwiller magnetic phase diagram of the cuprates},\ }\href
	{https://link.aps.org/doi/10.1103/PhysRevB.81.014509} {\bibfield  {journal} {\bibinfo
		{journal} {Phys. Rev. B}\ }\textbf {\bibinfo {volume} {81}},\ \bibinfo
	{pages} {014509} (\bibinfo {year} {2010}{\natexlab{b}})}\BibitemShut
	{NoStop}%
\bibitem [{\citenamefont {Gor'kov}\ and\ \citenamefont
  {Rashba}(2001)}]{gor2001}%
  \BibitemOpen
  \bibfield  {author} {\bibinfo {author} {\bibfnamefont {L.~P.}\ \bibnamefont
  {Gor'kov}}\ and\ \bibinfo {author} {\bibfnamefont {E.~I.}\ \bibnamefont
  {Rashba}},\ }\bibfield  {title} {\bibinfo {title} {Superconducting 2d system
  with lifted spin degeneracy: mixed singlet-triplet state},\ }\href@noop {}
  {\bibfield  {journal} {\bibinfo  {journal} {Physical Review Letters}\
  }\textbf {\bibinfo {volume} {87}},\ \bibinfo {pages} {037004} (\bibinfo
  {year} {2001})}\BibitemShut {NoStop}%
\bibitem [{\citenamefont {Qi}\ \emph {et~al.}(2010)\citenamefont {Qi},
  \citenamefont {Hughes},\ and\ \citenamefont {Zhang}}]{qi2010}%
  \BibitemOpen
  \bibfield  {author} {\bibinfo {author} {\bibfnamefont {X.-L.}\ \bibnamefont
  {Qi}}, \bibinfo {author} {\bibfnamefont {T.~L.}\ \bibnamefont {Hughes}},\
  and\ \bibinfo {author} {\bibfnamefont {S.-C.}\ \bibnamefont {Zhang}},\
  }\bibfield  {title} {\bibinfo {title} {Topological invariants for the fermi
  surface of a time-reversal-invariant superconductor},\ }\href@noop {}
  {\bibfield  {journal} {\bibinfo  {journal} {Physical Review B—Condensed
  Matter and Materials Physics}\ }\textbf {\bibinfo {volume} {81}},\ \bibinfo
  {pages} {134508} (\bibinfo {year} {2010})}\BibitemShut {NoStop}%
\bibitem [{\citenamefont {Yu}\ \emph {et~al.}(2011{\natexlab{a}})\citenamefont
  {Yu}, \citenamefont {Qi}, \citenamefont {Bernevig}, \citenamefont {Fang},\
  and\ \citenamefont {Dai}}]{yu2011}%
  \BibitemOpen
  \bibfield  {author} {\bibinfo {author} {\bibfnamefont {R.}~\bibnamefont
  {Yu}}, \bibinfo {author} {\bibfnamefont {X.~L.}\ \bibnamefont {Qi}}, \bibinfo
  {author} {\bibfnamefont {A.}~\bibnamefont {Bernevig}}, \bibinfo {author}
  {\bibfnamefont {Z.}~\bibnamefont {Fang}},\ and\ \bibinfo {author}
  {\bibfnamefont {X.}~\bibnamefont {Dai}},\ }\bibfield  {title} {\bibinfo
  {title} {Equivalent expression of z 2 topological invariant for band
  insulators using the non-abelian berry connection},\ }\href@noop {}
  {\bibfield  {journal} {\bibinfo  {journal} {Physical Review B—Condensed
  Matter and Materials Physics}\ }\textbf {\bibinfo {volume} {84}},\ \bibinfo
  {pages} {075119} (\bibinfo {year} {2011}{\natexlab{a}})}\BibitemShut
  {NoStop}%
\bibitem [{\citenamefont {Hao}\ \emph {et~al.}(2008)\citenamefont {Hao},
  \citenamefont {Zhang}, \citenamefont {Wang}, \citenamefont {Zhang},\ and\
  \citenamefont {Wang}}]{hao2008}%
  \BibitemOpen
  \bibfield  {author} {\bibinfo {author} {\bibfnamefont {N.}~\bibnamefont
  {Hao}}, \bibinfo {author} {\bibfnamefont {P.}~\bibnamefont {Zhang}}, \bibinfo
  {author} {\bibfnamefont {Z.}~\bibnamefont {Wang}}, \bibinfo {author}
  {\bibfnamefont {W.}~\bibnamefont {Zhang}},\ and\ \bibinfo {author}
  {\bibfnamefont {Y.}~\bibnamefont {Wang}},\ }\bibfield  {title} {\bibinfo
  {title} {Topological edge states and quantum hall effect in the haldane
  model},\ }\href@noop {} {\bibfield  {journal} {\bibinfo  {journal} {Physical
  Review B—Condensed Matter and Materials Physics}\ }\textbf {\bibinfo
  {volume} {78}},\ \bibinfo {pages} {075438} (\bibinfo {year}
  {2008})}\BibitemShut {NoStop}%
\bibitem [{\citenamefont {Hao}\ \emph {et~al.}(2011)\citenamefont {Hao},
  \citenamefont {Zhang},\ and\ \citenamefont {Wang}}]{hao2011}%
  \BibitemOpen
  \bibfield  {author} {\bibinfo {author} {\bibfnamefont {N.}~\bibnamefont
  {Hao}}, \bibinfo {author} {\bibfnamefont {P.}~\bibnamefont {Zhang}},\ and\
  \bibinfo {author} {\bibfnamefont {Y.}~\bibnamefont {Wang}},\ }\bibfield
  {title} {\bibinfo {title} {Topological phases and fractional excitations of
  the exciton condensate in a special class of bilayer systems},\ }\href@noop
  {} {\bibfield  {journal} {\bibinfo  {journal} {Physical Review B—Condensed
  Matter and Materials Physics}\ }\textbf {\bibinfo {volume} {84}},\ \bibinfo
  {pages} {155447} (\bibinfo {year} {2011})}\BibitemShut {NoStop}%
\bibitem [{\citenamefont {Sato}\ \emph {et~al.}(2011)\citenamefont {Sato},
  \citenamefont {Tanaka}, \citenamefont {Yada},\ and\ \citenamefont
  {Yokoyama}}]{sato20111}%
  \BibitemOpen
  \bibfield  {author} {\bibinfo {author} {\bibfnamefont {M.}~\bibnamefont
  {Sato}}, \bibinfo {author} {\bibfnamefont {Y.}~\bibnamefont {Tanaka}},
  \bibinfo {author} {\bibfnamefont {K.}~\bibnamefont {Yada}},\ and\ \bibinfo
  {author} {\bibfnamefont {T.}~\bibnamefont {Yokoyama}},\ }\bibfield  {title}
  {\bibinfo {title} {Topology of andreev bound states with flat dispersion},\
  }\href@noop {} {\bibfield  {journal} {\bibinfo  {journal} {Physical Review
  B—Condensed Matter and Materials Physics}\ }\textbf {\bibinfo {volume}
  {83}},\ \bibinfo {pages} {224511} (\bibinfo {year} {2011})}\BibitemShut
  {NoStop}%
\bibitem [{\citenamefont {Bednorz}\ and\ \citenamefont
  {M{\"u}ller}(1986)}]{bednorz1986}%
  \BibitemOpen
  \bibfield  {author} {\bibinfo {author} {\bibfnamefont {J.~G.}\ \bibnamefont
  {Bednorz}}\ and\ \bibinfo {author} {\bibfnamefont {K.~A.}\ \bibnamefont
  {M{\"u}ller}},\ }\bibfield  {title} {\bibinfo {title} {Possible high t c
  superconductivity in the ba- la- cu- o system},\ }\href@noop {} {\bibfield
  {journal} {\bibinfo  {journal} {Zeitschrift f{\"u}r Physik B Condensed
  Matter}\ }\textbf {\bibinfo {volume} {64}},\ \bibinfo {pages} {189} (\bibinfo
  {year} {1986})}\BibitemShut {NoStop}%
\bibitem [{\citenamefont {Kamihara}\ \emph {et~al.}(2008)\citenamefont
  {Kamihara}, \citenamefont {Watanabe}, \citenamefont {Hirano},\ and\
  \citenamefont {Hosono}}]{kamihara2008}%
  \BibitemOpen
  \bibfield  {author} {\bibinfo {author} {\bibfnamefont {Y.}~\bibnamefont
  {Kamihara}}, \bibinfo {author} {\bibfnamefont {T.}~\bibnamefont {Watanabe}},
  \bibinfo {author} {\bibfnamefont {M.}~\bibnamefont {Hirano}},\ and\ \bibinfo
  {author} {\bibfnamefont {H.}~\bibnamefont {Hosono}},\ }\bibfield  {title}
  {\bibinfo {title} {Iron-based layered superconductor la [o1-x f x] feas (x=
  0.05- 0.12) with t c= 26 k},\ }\href@noop {} {\bibfield  {journal} {\bibinfo
  {journal} {Journal of the American Chemical Society}\ }\textbf {\bibinfo
  {volume} {130}},\ \bibinfo {pages} {3296} (\bibinfo {year}
  {2008})}\BibitemShut {NoStop}%
\bibitem [{\citenamefont {Sun}\ \emph {et~al.}(2023)\citenamefont {Sun},
  \citenamefont {Huo}, \citenamefont {Hu}, \citenamefont {Li}, \citenamefont
  {Liu}, \citenamefont {Han}, \citenamefont {Tang}, \citenamefont {Mao},
  \citenamefont {Yang}, \citenamefont {Wang} \emph {et~al.}}]{sun2023}%
  \BibitemOpen
  \bibfield  {author} {\bibinfo {author} {\bibfnamefont {H.}~\bibnamefont
  {Sun}}, \bibinfo {author} {\bibfnamefont {M.}~\bibnamefont {Huo}}, \bibinfo
  {author} {\bibfnamefont {X.}~\bibnamefont {Hu}}, \bibinfo {author}
  {\bibfnamefont {J.}~\bibnamefont {Li}}, \bibinfo {author} {\bibfnamefont
  {Z.}~\bibnamefont {Liu}}, \bibinfo {author} {\bibfnamefont {Y.}~\bibnamefont
  {Han}}, \bibinfo {author} {\bibfnamefont {L.}~\bibnamefont {Tang}}, \bibinfo
  {author} {\bibfnamefont {Z.}~\bibnamefont {Mao}}, \bibinfo {author}
  {\bibfnamefont {P.}~\bibnamefont {Yang}}, \bibinfo {author} {\bibfnamefont
  {B.}~\bibnamefont {Wang}}, \emph {et~al.},\ }\bibfield  {title} {\bibinfo
  {title} {Signatures of superconductivity near 80 k in a nickelate under high
  pressure},\ }\href@noop {} {\bibfield  {journal} {\bibinfo  {journal}
  {Nature}\ }\textbf {\bibinfo {volume} {621}},\ \bibinfo {pages} {493}
  (\bibinfo {year} {2023})}\BibitemShut {NoStop}%
\bibitem [{\citenamefont {Yu}\ \emph {et~al.}(2011{\natexlab{b}})\citenamefont
  {Yu}, \citenamefont {Qi}, \citenamefont {Bernevig}, \citenamefont {Fang},\
  and\ \citenamefont {Dai}}]{PhysRevB.84.075119}%
  \BibitemOpen
  \bibfield  {author} {\bibinfo {author} {\bibfnamefont {R.}~\bibnamefont
  {Yu}}, \bibinfo {author} {\bibfnamefont {X.~L.}\ \bibnamefont {Qi}}, \bibinfo
  {author} {\bibfnamefont {A.}~\bibnamefont {Bernevig}}, \bibinfo {author}
  {\bibfnamefont {Z.}~\bibnamefont {Fang}},\ and\ \bibinfo {author}
  {\bibfnamefont {X.}~\bibnamefont {Dai}},\ }\bibfield  {title} {\bibinfo
  {title} {Equivalent expression of ${\mathbb{z}}_{2}$ topological invariant
  for band insulators using the non-abelian berry connection},\ }\href
  {https://doi.org/10.1103/PhysRevB.84.075119} {\bibfield  {journal} {\bibinfo
  {journal} {Phys. Rev. B}\ }\textbf {\bibinfo {volume} {84}},\ \bibinfo
  {pages} {075119} (\bibinfo {year} {2011}{\natexlab{b}})}\BibitemShut
  {NoStop}%
\end{thebibliography}
%
\end{document}